\documentclass[useAMS,twocolumn,usenatbib,fleqn]{mn2e} \voffset=-0.8in

\usepackage{etoolbox} \usepackage{ragged2e} \usepackage{xcolor} \usepackage{fontawesome} \usepackage{amssymb}

\usepackage[hidelinks,draft]{hyperref} \usepackage{graphicx} \usepackage{times} \usepackage{amsmath} \DeclareMathOperator{\Tr}{Tr} \DeclareMathOperator{\diag}{diag}

  \def\apj{ApJ} \def\apjs{ApJS}  \def\mnras{MNRAS}  \def\prd{PhysRevD}    \def\apss{APSS} \def\aap{A\&A}  \def\na{NA}

\usepackage{pifont}  

\newcommand{\GG}[1]{}

\usepackage{natbib} 

\begin{document}

\title[$N$-body integration with {\sc exp}] {{\sc exp}: $N$-body integration using basis function expansions}

\author[Petersen, Weinberg, \& Katz] {Michael~S.~Petersen,$^{1}$\thanks{michael.petersen@roe.ac.uk} Martin~D.~Weinberg,$^2$ Neal~Katz$^2$\\ $^1$Institute for Astronomy, University of Edinburgh, Royal Observatory, Blackford Hill, Edinburgh EH9 3HJ, UK \\ $^2$Department of Astronomy, University of Massachusetts at Amherst, 710 N. Pleasant St., Amherst, MA 01003}

\maketitle

\begin{abstract} 
We present the \(N\)-body simulation techniques in {\sc exp}. {\sc exp} uses empirically-chosen basis functions to expand the potential field of an ensemble of particles. Unlike other basis function expansions, the derived basis functions are adapted to an input mass distribution, enabling accurate expansion of highly non-spherical objects, such as galactic discs. We measure the force accuracy in three models, one based on a spherical or aspherical halo, one based on an exponential disc, and one based on a bar-based disc model. We find that {\sc exp} is as accurate as a direct-summation or tree-based calculation, and in some ways is better, while being considerably less computationally intensive. We discuss optimising the computation of the basis function representation. We also detail numerical improvements for performing orbit integrations, including timesteps.
\end{abstract}


\section{Introduction} \label{sec:introduction}

The \(N\)-body technique of simulating dynamical systems computes the forces at the location of each of \(N\) particles in a simulation from the other \(N-1\) particles. Generally speaking, the forces are then applied to each particle, advancing the system forward in time, and the process of computing the forces is repeated. Unfortunately for those seeking to model Milky Way-like galaxies, the wide range of spatial scales (from sub-pc to hundreds of kpcs), and the wide range of temporal scales (from years to Gyrs) makes the computational requirements for computing the forces by the direct interaction of particles presently infeasible. Therefore, one seeks different, expedient techniques to compute the potential fields and calculate the forces on each individual particle.

One such technique is the basis-function expansion (BFE) technique, where one uses appropriately chosen biorthogonal basis functions that solve the Poisson equation using separable azimuthal harmonics, \(m\).  One may then inexpensively obtain the forces at any point in space as follows. A biorthogonal system is a pair of indexed families of functional vectors in some topological functional vector space such that the inner product of the pair is the Kronecker delta. The BFE method constructs a biorthogonal system \(\{\phi_\mu(\mathbf{x}), d_\mu(\mathbf{x})\)\}\ such that \(\nabla^2 \phi_\mu = 4\pi G d_\mu\) and \(\int d\mathbf{x}\, \phi_\mu(\mathbf{x}) d_\nu(\mathbf{x}) = 4\pi G\delta_{\mu\nu},\) where \(\delta_{\mu\,\nu}\) is the Kronecker delta and the individual functions are denoted by Greek letters. Consider the density of some target distribution, \(\rho (\mathbf{x})\). Approximations for the density and potential fields when using $M$ total basis functions (denoted by \(\check\cdot\)) are 
\begin{equation} \check{\rho}(\mathbf{x}) = \sum_{\mu=1}^M a_\mu d_\mu(\mathbf{x}) \label{eq:approxrho}
\end{equation} and \begin{equation}
\check{\Phi}(\mathbf{x}) = \sum_{\mu=1}^M a_\mu \phi_\mu(\mathbf{x}) \end{equation} 
where the amplitudes of the {\it coefficients}, \(a_\mu\), of each of the $\mu$ functions are given by \begin{equation} a_\mu = \int d\mathbf{x}\, \rho(\mathbf{x}) \phi_\mu(\mathbf{x}). \label{eq:amplitude} \end{equation}

In the case of an \(N\)-body simulation, let the density of our particle distribution of \(N\) points with individual masses \(m_i\) be described by \begin{equation} \rho(\mathbf{x}) = \sum_{i=1}^N m_i \delta\left(\mathbf{x} - \mathbf{x_i}\right), \end{equation} 
making the coefficients that approximate the potential \begin{equation} \hat{a}_\mu = \frac{1}{N} \sum_{i=1}^N \phi_\mu(\mathbf{x}_i). \label{eq:coefficients} \end{equation}
We will hereafter denote quantities computed with an ensemble of discrete particles using \(\hat\cdot\).

Generally, any technique making use of these properties is called a BFE method \citep{cluttonbrock72, cluttonbrock73, kalnajs76, hernquist92,earn96, weinberg99}. BFE methods have many features that make them ideal for studying disturbances to equilibrium stellar discs and dark matter halos (or other spherical systems). For simulations using BFE methods, harmonic function analysis decomposes a distribution into linearly-summable functions that resemble expected evolutionary scenarios in disc galaxy evolution. 

The $N$-body code {\sc exp} has been discussed elsewhere in piecemeal chunks \citep{weinberg99,weinberg02,holleyb05,weinberg07a,weinberg07b,choi07,choi09,petersen16a}. We collect and update the various algorithms that are now part of {\sc exp} in this paper. In this work, we describe the primary technique unique to {\sc exp}: adaptive bases constructed to reproduce equilibria with one or at most several basis functions. This yields a rapid convergence in the expansion series. We have recently used {\sc exp} to study bar formation in a stellar halo embedded in the dark matter halo \citep{petersen19,petersen21}, but the techniques are generically applicable. By combining multiple bases for different scales and geometries, we can decompose a galaxy model based on the geometry and symmetry of the different components. In the dark matter halo and stellar disc case, this corresponds to two separate sets of basis functions, one for each component. Both the halo and disc use three-dimensional orthogonal functions to represent the potential. The basis functions then enable a quick and straightforward reconstruction of the potential at any time in the simulation, for any arbitrary combination of particles. Tracking the amplitudes of the basis functions through time is the primary investigative tool used when analysing basis function expansion simulations. The amplitudes efficiently summarise the degree and nature of asymmetries in the potential.

In the next section, we describe how to select the basis functions for a given mass distribution (Section~\ref{section:bfe}). We describe the theory that underpins the basis selection (Section~\ref{subsec:conditioning}) and describe the two relevant examples for a halo-disc system: a spherical expansion (Section~\ref{subsec:spherical}) and a cylindrical expansion (Section~\ref{subsec:cylindrical}). In Section~\ref{sec:validation}, we also test the accuracy of the BFE forces for the case of a spherical and aspherical halo, an exponential disc, and a bar model. We then turn to evolving the phase space in time in Section~\ref{sec:expevolution}, including integration, coefficient calculation, and timesteps. We summarise in Section~\ref{sec:expsummary}.

\section{Basis Function Expansion} \label{section:bfe}

The first step in constructing a BFE simulation is to select the basis with which the forces on each particle will be computed. We detail the procedure for conditioning the basis in Section~\ref{subsec:conditioning}, including the derivation of new orthogonal bases that best represent a specified input density distribution, such as an axisymmetric disc.

As described in equations~(\ref{eq:approxrho})-(\ref{eq:coefficients}), the BFE method computes the gravitational potential by projecting particles onto a set of biorthogonal basis functions that satisfy the Poisson equation. The Poisson equation is separable in any conic coordinate system. Each equation in the separation has a Sturm-Louiville (SL) form. The SL equation describes many physical systems, and may be written as: \begin{equation} \frac{d}{dx}\left[p(x)\frac{du(x)}{dx}\right] - q(x)u(x) = \lambda w(x) u(x) \label{eq:sle} \end{equation} where $\lambda$ is a constant, and $w(x)>0$ is a weighting function. The function $u(x)$ is the unknown function and $p(x)$, $q(x)$, and $w(x)$ are parameter functions. The eigenfunctions $u_\mu$ of the SL equation form a complete basis set with eigenvalues $\lambda_\mu$ where the formally infinite series in $\mu$ may be truncated \citep{courant89}. If one writes the Poisson equation for Cartesian coordinates in the form of equation~(\ref{eq:sle}), then $p(x)=1,q(x)=0,w(x)=1$, and the solution is sines and cosines. For the Poisson equation in spherical coordinates, the solution is spherical harmonics for the angular coordinates and Bessel's equation for the radial equation.  However, by changing the weighting function $w(x)$, one may derive an infinite number of radial bases. {\sc exp} exploits this flexibility to select a weighting function $w(x)$ such that the unperturbed potential may be represented by a single term (see Section \ref{subsec:spherical}). We refer to this as {\sl adaptive basis conditioning}. Because one designs the lowest-order term to match the initial profile, we are able to accurately represent the potential on chosen scales using only a small number of higher-order terms. Provided that the structure still resembles the initial model, even after some evolution one may represent the potential to high spatial accuracy.

\begin{figure} \centering \includegraphics[width=3.3in]{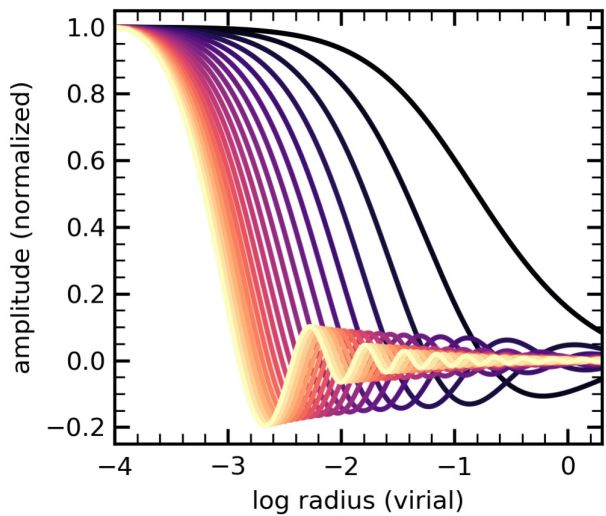} \caption{\label{fig:ufuncs} Eigenfunction amplitude as a function of radius for an example spherical basis. We show the first 20 eigenfunctions, from low (dark colour) to high (light colour). } \end{figure}

\begin{figure*} \centering \includegraphics[width=6.7in]{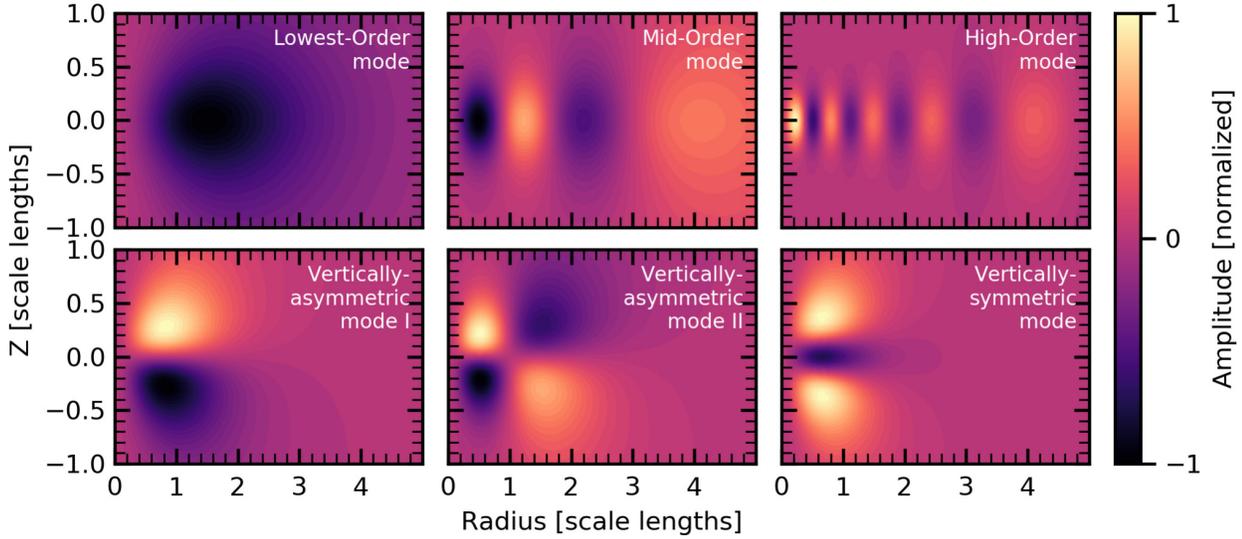} \caption{\label{fig:discrz} An example of different potential basis terms $\phi_\mu$ for the disc, as a slice at azimuth $\phi=0$ in radius--vertical space. These modes are all found in a disc basis realisation, but may or may not be present in an individual realisation depending upon the radial truncation order $n_{\rm max}$. } \end{figure*}

\subsection{Adaptive basis conditioning} \label{subsec:conditioning}

Let the contribution of some particle at position \(\mathbf{x}\) to the BFE coefficients for some fiducial basis be given by the vector \(\mathbf{c} = \left(\phi_1(\mathbf{x}), \phi_2(\mathbf{x}), \ldots, \phi_n(\mathbf{x})\right)\) where \(\phi_\mu\) are the basis functions.  Denote the ensemble average over the particle distribution by \(\langle\cdot\rangle\).  For example, \(\langle\mathbf{c}\rangle\) is the expected BFE coefficient vector, normalised by mass. For a convergent expansion, the elements of \(\langle\mathbf{c}\rangle\) will decrease to zero with \(\mu\). If the basis is not optimal for the target distribution, it may converge slowly. However, one may find a new basis that converges more quickly to the desired density by assuming the existence of some unitary transformation \(\mathbfss{U}\) to a new basis \(\mathbf{b} = \mathbfss{U}\cdot\mathbf{c}\).  That is, the columns of \(\mathbfss{U}\) are new orthogonal functions (vectors) with \(\mathbfss{U}^\intercal\cdot\mathbfss{U} = \mathbfss{1}\).  Each member of the function space spanned by our initial biorthogonal basis is a solution of the Poisson equation. The linearity of the Poisson equation implies that the new functions represented by the unitary transformation are also a solution.  Completeness of the new basis may be argued similarly. The basis transformed by \(\mathbfss{U}\) is an equivalent basis.  The derivation of such an optimal transformation is closely related to mean integrated square error (MISE) extremisation under truncation. In this section, we derive the transformation to this new basis.

We posit the existence of a truncated basis of arbitrarily lower rank to represent the content of \(\langle\mathbf{c}\rangle\). Our truncated basis takes the form \(\tilde{\mathbf{c}} = \mathbfss{U}\cdot\tilde{\mathbf{b}}\) where \(\tilde{\mathbf{b}} = \left(b_1, b_2,\ldots, b_M, 0, 0, \ldots\right)\) with \(M\le n\) enforcing the truncation.\footnote{Intuitively, it may be helpful to think of a \(M\ll n\) with \(n\) larger than any practically computatable value. However, the arguments here do not require this ordering.}  Summed over all particles, the mean-squared error for the truncation relative to the original representation is
\begin{equation}
  \mathcal{E}(M) = \langle\left(\mathbf{c} -
  \tilde{\mathbf{c}}\right)^2\rangle = \langle c^2\rangle -
  2\langle\mathbf{c}\cdot\tilde{\mathbf{c}}\rangle + \langle\tilde{c}^2\rangle.
  \label{eq:expectation}
\end{equation}
The first term in the last equality of equation~(\ref{eq:expectation}) is independent of \(M\), but the second and third terms are not. We may rewrite the second term as
\begin{equation}
\langle\mathbf{c}\cdot\tilde{\mathbf{c}}\rangle =
\langle\mathbf{c}\cdot\mathbfss{U}^\intercal\cdot\tilde{\mathbf{b}}\rangle =
\langle\mathbfss{U}^\intercal\cdot\mathbf{b}\cdot\mathbfss{U}^\intercal\cdot\tilde{\mathbf{b}}\rangle
= \langle\mathbf{b}\cdot\tilde{\mathbf{b}}\rangle = \langle
  \tilde{b}^2\rangle.
\end{equation}
By a similar computation, the third term \(\langle\tilde{c}^2\rangle=\langle\tilde{b}^2\rangle\).  We can then rewrite equation~(\ref{eq:expectation}) as
\begin{equation}
  \mathcal{E}(M) = \langle c^2\rangle - \langle \tilde{b}^2\rangle.
\end{equation}
This last equation implies that \(\mathcal{E}(M)\) will have a minimum when \(\langle \tilde{b}^2\rangle\) is maximised, subject to the orthogonality constraint. We can write this requirement as the maximum of the function
\begin{equation}
  \mathcal{F}(M) =
  \left\langle\left(\mathbfss{U}\cdot\mathbf{c}\right)^2\right\rangle -
\Tr \left[\mathbfss{L}\cdot\left(\mathbfss{U}^\intercal\mathbfss{U} -
  \mathbfss{1}\right) \right]
  \label{eq:tomaximise}
\end{equation}
where \(\mathbfss{L} = \diag\{\lambda_1,\ldots,\lambda_M\}\) is the diagonal matrix of Lagrange multipliers and \(\mathbfss{1}\) is the identity matrix.  The extremum of \(\mathcal{F}(M)\) is the solution to
\begin{equation}
  0 = \frac{\partial \mathcal{F}(M)}{\partial\mathbfss{U}}
  = 2\mathbfss{U}\cdot\langle\mathbf{c}\otimes\mathbf{c}\rangle -
  2\mathbfss{L}\cdot\mathbfss{U}
  \label{eq:matrixequation}
\end{equation}
where \(\otimes\) denotes the vector outer product.
Equation~(\ref{eq:matrixequation}) is the matrix equation
\begin{equation}
  \mathbfss{D}\cdot\mathbfss{U} = \mathbfss{U}\cdot\mathbfss{L}
  \label{eq:usoln}
\end{equation}
where \(\mathbfss{D} = \langle\mathbf{c}\otimes\mathbf{c}\rangle\).
The elements of \(\mathbfss{D}\) are
\begin{equation}
  \mathbfss{D}_{\mu\nu} = \langle\phi_\mu(\mathbf{x}) \phi_\nu(\mathbf{x})\rangle.
\end{equation}
We may estimate \(\mathbfss{D}_{\mu\nu}\) for a sample \(N\) particles with mass \(m\) as
\begin{equation}
  \mathbfss{D}_{\mu\nu} = \frac{m}{M}\sum_{i=1}^N\phi_\mu(\mathbf{x}_i) \phi_\nu(\mathbf{x}_i).
    \label{eq:withparticles}
\end{equation}
In the infinite particle limit, the ensemble average becomes
\begin{equation}
  \mathbfss{D}_{\mu\nu} = \frac{1}{M}\int d^3\mathbf{x} \rho(\mathbf{x}) \phi_\mu(\mathbf{x}) \phi_\nu(\mathbf{x}).
  \label{eq:withoutparticles}
\end{equation}
The mass normalisation does not affect the solution and may be discarded in practice. One may either condition on the particle distribution (equation~\ref{eq:withparticles}) or, in the infinite particle limit, using analytic functions (equation~\ref{eq:withoutparticles}).

The matrix \(\mathbfss{D}_{\mu\nu}\) describes which terms, \(a_\mu\), contribute the most to the gravitational energy.  The orthonormal basis that diagonalises \(\mathbfss{D}\) has the target density as its lowest-order basis function. Because \(\mathbfss{D}\) is symmetric and positive definite, all the eigenvalues are positive. The term with the largest eigenvalue describes the majority of the correlated contribution, and so on for the second largest eigenvalue, etc. {\sc exp} performs this diagonalisation using singular value decomposition (SVD) and the singular matrices (now mutual transposes owing to symmetry) describe a rotation of the original basis into the uncorrelated basis. 

The desired unitary transformation that best represents the gravitational field with \(M\le n\) follows from the singular value decomposition of \(\mathbfss{D}\) (equation~\ref{eq:usoln}). The eigenfunction corresponding to the largest singular value in \(\mathbfss{D}\) describes a new basis function with the largest contribution to the gravitational energy, the next eigenfunction/eigenvalue pair describes the next largest contribution, and so on for each successive eigenfunction/eigenvalue. Each basis function is uncorrelated with the others by construction. The new basis functions optimally approximate the true distribution from the spherical-harmonic expansion in the original basis in the sense that the largest amount of gravitational field energy is contained in the smallest number of terms; the SVD solution provides the linear solution with minimum length \citep{strang06}.  One might call this optimal in the \emph{least-squares sense} \citep{weinberg96}. The new coefficient vector is related to the original coefficient vector by the orthogonal transformation defined by the singular vectors of the SVD. Since the transformation and the Poisson equation are linear, the new eigenfunctions are also biorthogonal.  

We call the process of performing the singular value decomposition of \(\mathbfss{D}\) to derive the empirical orthogonal functions that best describe the target density distribution \emph{adaptive basis conditioning}. In our first example, the spherical basis representing the halo, we may use a direct solution of the SL equation. In the second example, the cylindrical basis, we use an existing spherical basis and construct \(\mathbfss{D}\) to define a new orthogonal linear combination of basis functions that best matches a stellar disc. In this case, the resulting unitary transformation provides a representation of the disc basis based on the input spherical basis. This \emph{conditions} the original basis to vary only in the vicinity of non-negligible disc density (see section~\ref{subsec:cylindrical} for details). For accuracy, the required number of spherical basis terms is large. However, the transformation only needs to be performed once, and the resulting disc basis functions may be tabulated for further use.

\begin{figure} \centering \includegraphics[width=3.5in]{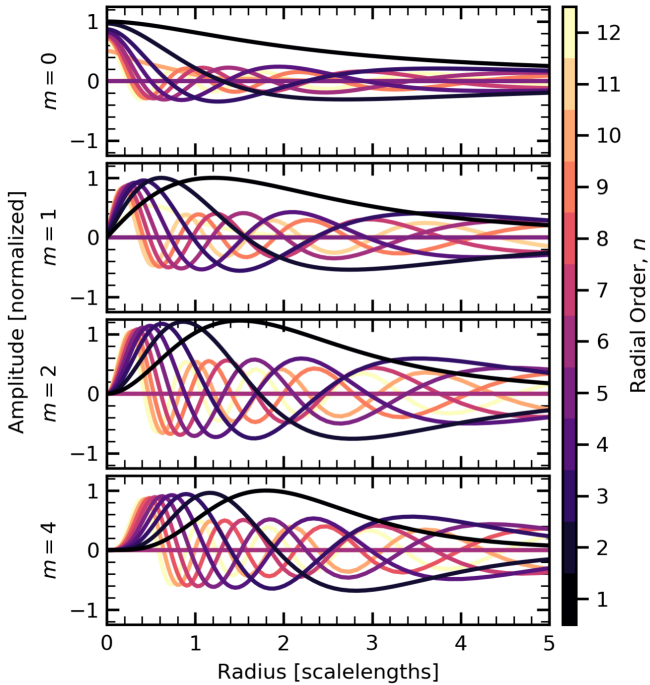} \caption{\label{fig:disc_amplitudes} Amplitude of the disc basis functions as a function of radius in scale lengths for four azimuthal ($m$) orders, when $z=0$ (in-plane) and azimuth $\phi=0$. The first 12 radial ($n$) orders are shown.} \end{figure}

\subsection{Example 1: A Spherical Basis} \label{subsec:spherical}

A relatively straightforward example of an adaptive basis is a spherically-symmetric density profile, such as an NFW halo profile. For an expansion in spherical harmonics, the Poisson equation separates into angular and radial equations. The solution to the angular equations are spherical harmonics. For the radial equation, following the notation of equation~(\ref{eq:sle}), we choose a weighting function $w(r)\propto \rho_0(r)\Phi_0(r)$ as described in \citet{weinberg99}, where $\rho_0(r),\Phi_0(r)$ is the unperturbed model. Substituting $\Phi(r)=\Phi_0(r)u(r)$ into the Poisson equation, it immediately follows that the functions in equation~(\ref{eq:sle}), with $x=r$, become: \begin{equation} p(r) = r^2\Phi_0^2(r), \end{equation} \begin{equation} q(r) = \left[l(l+1)\Phi_0(r) - \nabla_r^2\Psi_0(r)r^2\right]\Phi_0(r), \end{equation} and \begin{equation} w(r) = -4\pi G r^2\Phi_0(r)\rho_0(r). \end{equation} Equation~(\ref{eq:sle}) then yields a series of eigenfunctions $u_n(r)$ with potentials $\phi_n\propto\Phi_0(r)u_n(r)$. Each eigenfunction $u_n$ then corresponds to a solution of equation~(\ref{eq:sle}). Examples of eigenfunctions \(u_n\) are shown in Figure~\ref{fig:ufuncs}. The lowest-order eigenfunction with \(\lambda=1\) is \(u_1(r)=1\) by construction.  Each successive $u_n(r)$ with $n>1$ has an additional radial node.  The prefactor $\Phi_0(r)$ ensures that the basis resembles the target model. The radial boundary conditions are straightforward to apply at the origin and at infinity. See \citet{weinberg99} for more details. In spherical coordinates $(r,\theta,\phi)$, the general index $\mu$ is the triple ($l,m,n$), where $l$ and $m$ are standard spherical harmonic angular indices and $n$ is the radial index. Replacing $u_\mu \to u^{lm}_n$ then specifies the eigenfunction per each set of angular coordinates $(l,m)$. Each $\phi_\mu$ term in the total halo potential is then given by $\phi_{\mu}(\mathbf{x}) \to \phi^{lm}_{n}(\mathbf{x}) = \Phi_0(r)u^{lm}_n(r)Y_{lm}(\theta,\phi)$.

The angular structure is set by a combination of the radial potential functions, as tabulated from the eigenfunctions of the SL equation, and spherical harmonics. The angular basis functions for each spherical harmonic\footnote{In practice, the radial basis functions are the same for each $l$ order, independent of $m$, reducing the number of eigenfunctions that must be tabulated.} are chosen as the first $n$ eigenfunctions sorted by decreasing eigenvalue of the matrix \(\mathbfss{U}\) (see eqs. \ref{eq:withparticles},  \ref{eq:withoutparticles}). The coefficients for each eigenfunction in $r$ and $\cos\theta$ have cosine and sine components that correspond to the analogous Fourier terms. The sine and cosine terms of each azimuthal order give the phase angle of the harmonic and can be used, for example, to calculate the pattern speed in an evolving system.

\subsection{Example 2: A Cylindrical Basis} \label{subsec:cylindrical}

To represent a disc, we want a basis whose vertical dimension matches that of the input density. Although one can construct a cylindrical disc basis from the eigenfunctions of the Laplacian as in the spherical case \citep[e.g.][]{earn96}, the boundary conditions in cylindrical coordinates ($R$, $\phi$, $z$) make the basis hard to implement. To circumvent this complexity, our method starts with a three-dimensional spherical basis with a high harmonic order, e.g. $l_{\rm initial},m_{\rm initial}\le64$, and defines a unitary vector-space transformation to a new basis that best represents the target disc density using a lower number of harmonic orders, e.g. $m_{\rm final}=\{6,8,12\}$, where the choice of $m_{\rm final}$ is dictated by the problem at hand.

The three dimensional functions in spherical coordinates $(r,\theta,\phi)$ that represent the target disc density are then transformed to cylindrical coordinates $(R,\phi,z)$. The spherical basis functions are all proportional to \(\cos(m\phi)\) or \(\sin(m\phi)\) so the cylindrical coordinates of the result disc functions have the same dependence. Therefore, we can tabulate their dependence uniquely in meridional cylindrical coordinates \((R,z)\) for each cosine or sine term. Then, each density element \(\rho(R, \phi, z)\,d^3x\) contributes \begin{equation} \frac{1}{4\pi G}u^{lm}_n (r)Y_{lm}(\theta,\phi)\rho(R, \phi, z)d^3 x \end{equation} to the expansion coefficient \(a_{\mu}\), or \begin{equation} a_{\mu} = \frac{1}{4\pi G}\int u^{lm}_n (r)Y_{lm}(\theta,\phi)\rho(R, \phi, z)d^3x \label{eq:cylanalytic}\end{equation} and \begin{equation} \hat{a}_{\mu}= \lim_{N\rightarrow\infty}\frac{1}{4\pi G}\sum_{i=1}^N m_i u^{lm}_n (r_i)Y_{lm}(\theta_i,\phi_i). \label{eq:cylparticle}\end{equation} The second equation shows the Monte Carlo approximation for \(N\) particles where \(\sum_i m_i = \int \rho(R,\phi, z)d^3x\).

We construct the matrix \(\mathbfss{D}\) (eqn. \ref{eq:withparticles}) given the density \(\rho(R, \phi, z)\) from the initial high harmonic order basis. The elements are products of $Y_{lm}u^{lm}_n$ and $Y_{l^\prime m}u^{l^\prime m}_{n^\prime}$ integrated over density. The eigenfunctions of the matrix \(\mathbfss{U}\) for the spherical-harmonic decomposition restricted to a particular azimuthal harmonic $m$ yield basis functions defined on the meridional plane $(R, z)$. We select the $n_{\rm max}$ most significant meridional eigenfunctions. Each of these basis functions vary both radially and vertically.  Because they are conditioned by a three-dimensional disc density, the lowest order functions look like the disc density.  As the index of the meridional basis function increases, the length scale of both radial and vertical variations decreases.  The symmetry of the original basis and the target density ensures that the basis functions have both even and odd parity with respect to the midplane. 

We condition the initial disc basis functions on the analytic disc density.  The resulting lowest-order potential density pair will be close to the analytic profile (eqns.~\ref{eq:cylanalytic} and \ref{eq:cylparticle}). For typical particle numbers $N$ in simulations, using the analytic disc density acts to reduce small-scale discreteness noise as compared to conditioning the basis function on the realised positions of the particles \citep{weinberg98}. We choose a spherical profile based on the deprojected disc density.

Examples in $(R,z)$ are shown in Figure~\ref{fig:discrz}. Figure~\ref{fig:disc_amplitudes} shows the in-plane \((z=0)\) behaviour of the meridional plane basis functions ($n$ orders) as a function of radius, for each harmonic subspace ($m$ orders). We show the four azimuthal harmonics that are most relevant for the evolution of a disc simulation, $m=0,1,2,4$, from top to bottom in the panels. For each harmonic, the lowest-order meridional index, $n=1$, has no nodes except at $R=0$. The number of nodes increases with order $n$. The nodes are interleaved for increasing meridional order. The increasing number of nodes means that the smallest meridional node always decreases in radius as the number of nodes increases.  The spacing of nodes gives an approximate value for the adaptive force scale length of the simulation. For example, the highest order $m=0$ meridional function ($n=12$) has a zero at $R=0.2R_d$, or 600 pc in a MW-like galaxy. Additionally, the meridional indices are interleaved between harmonic orders, such that $R_{\rm first node,m=2,n=1} \approx \frac{1}{2}\left(R_{\rm first node,m=1,n=1} + R_{\rm first node,m=1,n=2}\right)$. In practice, we select $n_{\rm max}$ to provide sensitivity to spatial scales of approximately 100 pc.  Note that the node spacing is not the same as spatial resolution in a standard particle code: the choice of order that determines the variational scale is not the same as spatial resolution in this method. For example, even though the highest $n$ would only imply a spatial variation with a scale of 100 pc, the basis resolves the analytic density down to 10 pc. Furthermore, the choice of a maximum meridional index removes or filters high spatial frequencies that result from particle noise.

\section{Validation} \label{sec:validation}

We now measure the accuracy of the expansions by comparing the expanded forces to a high-accuracy force estimate. We do so for the spherical (Section~\ref{subsec:spherical}) and cylindrical (Section~\ref{subsec:cylindrical}) bases by realising test distributions of particles. We also briefly discuss the significance of the coefficients (Section~\ref{subsec:coefficientsignificance}), deferring a detailed study to future work. We summarise the findings in Section~\ref{subsec:forcesummary}.

\subsection{Spherical Expansions}\label{subsec:spherical}

In a spherical model, the lowest-order function is defined to match the density and potential of the input spherical system. To evaluate the accuracy, we compare the expanded forces to true forces:
\begin{eqnarray}
\Delta_r(r) &= \frac{\left|F_{r,~{\rm expanded}}(r) -F_{R,~{\rm analytic}}(r)\right|}{F_{r,~{\rm analytic}}(r)} \label{eq:raderr_sph}\\ 
\Delta_\theta(r) &= \frac{\left|F_{\theta,~{\rm expanded}}(r) -F_{\theta,~{\rm analytic}}(r) \right|}{\sqrt{F_{r,~{\rm analytic}}^2+F_{\theta,~{\rm analytic}}^2+F_{\phi,~{\rm analytic}}^2}(r)} \label{eq:polarerr_sph}\\
\Delta_\phi(r) &= \frac{\left|F_{\phi,~{\rm expanded}}(r) -F_{\phi,~{\rm analytic}}(r) \right|}{\sqrt{F_{r,~{\rm analytic}}^2+F_{\theta,~{\rm analytic}}^2+F_{\phi,~{\rm analytic}}^2}(r)}. \label{eq:phierr_sph} \end{eqnarray}
where $F_r$ is the three-dimensional radial force, $F_\theta$ is the polar force, $F_\phi$ is the azimuthal force, and the subscripts `expanded' and `analytic' refer to the BFE forces and the exact forces, respectively. We normalise the polar and azimuthal forces by the total force to avoid zeros in the polar and azimuthal force that result from symmetries in the test density configurations. 

To put our BFE force errors in context, we also compute the same relative force errors for the tree-code gravity of {\sc gadget-2} \citep{springel05}. The relative forces are computed using equations~(\ref{eq:raderr_sph})--(\ref{eq:phierr_sph}), replacing the expanded forces with tree-gravity forces. As is always done for these types of gravity solvers, we soften the gravity and we choose a cubic spline kernel \citep{monaghan85,hernquist89}, with a softening length of $h=0.0011R_{\rm vir}$. As discussed in \citet{springel05}, the cubic spline has a potential equivalent to a point mass at zero lag as a Plummer softening when the kernel width $h=2.8\epsilon$, where $\epsilon$ is the standard Plummer softening length. We choose our softening length $h$ by downscaling the softening length in the simulations of \citet{donghia19} to match our particle number. We verified that this choice of softening length is near the minimum force error for softened gravity by testing values of $h$ a factor of two larger and smaller, finding that the median force errors increase in either direction. For {\sc gadget-2}, we input the softening length $\epsilon=0.0004R_{\rm vir}$.

Partitioning the particles into bins, we compute the mean, $\mu = \langle \Delta_{\{R,z,\phi\}} \rangle$, and the variance, $\sigma^2=\sigma(\Delta_{\{R,z,\phi\}})^2$, in each bin. To measure $\sigma(\Delta_{\{R,z,\phi\}})^2$, we compute the difference between the 16th and 84th percentile relative force values in each bin, to capture 2/3 of the distribution and to trim outliers. For the rest of the work, we report the root variance, \(\sigma\).

\begin{table*} \begin{tabular}{ccccccc} 
\hline 
\((l,n)\) & \(\mu_{\Delta_r}\) [\%] & \(\mu_{\Delta_r} (r<0.01R_{\rm vir})\) [\%] & \(\mu_{\Delta_\theta}\) [\%] & \(\mu_{\Delta_\theta} (r<0.01R_{\rm vir})\) [\%] & \(\mu_{\Delta_\phi}\) [\%] & \(\mu_{\Delta_\phi} (r<0.01R_{\rm vir})\) [\%]  \\ 
\hline 
(0,1) & 1.2e-2 & 1.3e-2 & 4.5e-7 & 3.8e-7 & 3.2e-7 & 2.9e-7 \\
(6,24) & 6.4e-2 & 8.0e-1 & 4.9e-4 & 3.5e-2 & 3.5e-4 & 3.5e-2 \\
\hline
{\sc gadget-2} &  2.4e-1 & 2.63 & 2.2e-3 & 3.9e-2 & 1.6e-3 & 3.0e-2 \\
\hline 
\end{tabular} 
\caption{Overall force accuracy measures for different force realisations in the near-ideal NFW case. \label{tab:forceerror}  } \end{table*}

\subsubsection{An NFW model} \label{subsubsec:basenfw}

As a first example, we generate basis functions for a idealised halo that resembles the expected Milky Way parameters. First, we specify a spherically-symmetric Navarro-Frank-White (NFW) dark matter halo radial profile \citep{navarro97}: \begin{equation} \rho_h(r) = \frac{\rho_0r_s^3}{\left(r+r_c\right)\left(r+r_s\right)^2} \label{eq:nfw} \end{equation} where $\rho_0$ is a normalisation set by the chosen mass, $r_s=0.083R_{\rm vir}$ is the scale radius, $R_{\rm vir}=1$ is the virial radius, and $r_c=0.0002R_{\rm vir}$ is a radius that sets the size of the core, i.e. where the density $\rho_h(r)$ becomes constant with radius. While the pure NFW halo has $r_c=0$, we choose a small value to avoid the formal divergence when $r_c=0$ and $r\to0$. We also include an error function at \(>2R_{\rm vir}\) to give a finite mass: \(\rho_{\rm halo, trunc}(r) = \rho_{\rm halo}(r)\left[\frac{1}{2}-\frac{1}{2}\left({\rm erf}\left[(r-r_{\rm trunc})/w_{\rm trunc}\right]\right)\right]\), where $r_{\rm trunc}=2R_{\rm vir}$ and $w_{\rm trunc}=0.3R_{\rm vir}$. We realise the particle locations by Eddington inversion \citep{binney08}. We refer to this model as the `near-ideal' case. 

We derive the corresponding empirical orthogonal basis functions, as in Section~\ref{subsec:spherical}. We compute coefficients for the particle representation by solving equation~(\ref{eq:coefficients}) using the derived basis functions. A typical simulation uses harmonics \(l_{\rm max}=6\) and radial order \(n_{\rm max}\le24\), where the choice of \(n_{\rm max}\) is set by the problem at hand. For the purposes of testing the basis accuracy, we retain \(n_{\rm max}=24\). We then use the coefficients to calculate the forces and compare with the analytic (exact) solutions.

As expected, the lowest order function \((l,n)=(0,1)\) is a near-perfect representation of the forces. As the distribution is initially spherical, we expect that the inclusion of any additional functions will decrease the overall accuracy. However, the additional terms must be included to resolve the subsequent evolution of the system. Therefore, while we report the mean of the force accuracy, \(\mu\), the root variance of the force accuracy, \(\sigma\), may be the more interesting quantity as it measures the `noise' in the expansion. Noise from discreteness maybe be physical (e.g. from subhaloes) and aphysical (e.g. on scales without a natural cause).

The vast majority of particles experience minimal force bias from noisy coefficients, as evidenced by a measure of the force bias for all particles, shown in Table~\ref{tab:forceerror} for the idealised single-term case and the typical \((l=6,n=24)\) combination. Even for particles inside of a disc radius, \(r<0.01R_{\rm vir}\), we find that the overall force bias and variance are small. For our canonical \((l,n)=(6,24)\) choice, the typical force errors for particles are less than 1\%.

In the next section, we turn to cases where the represented distribution does not perfectly match the basis.

\begin{figure} \centering \includegraphics[width=3.4in]{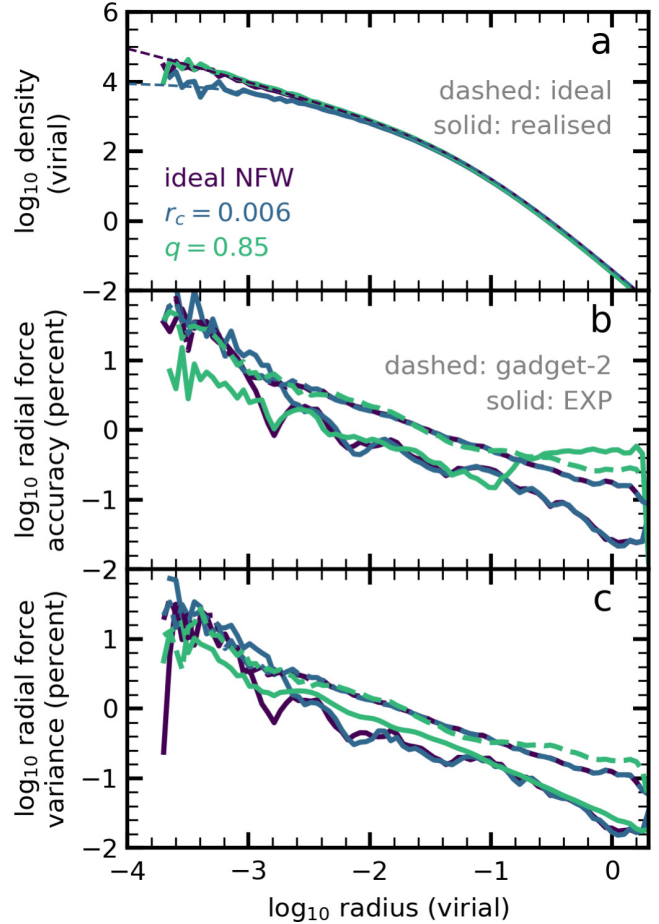} \caption{\label{fig:sph_accuracy_compare} Measurement of the logarithm of mean force accuracy (panel b) and the logarithm of the root variance of force accuracy (panel c) as a function of the logarithm of the radius for the three profiles discussed in the text (coded by colour, with spherically-averaged profiles shown in panel a). In panels b and c, we show the forces computed using {\sc gadget-2} with corresponding colour dashed lines for reference.} \end{figure}

\subsubsection{Non-ideal NFW models} \label{subsubsec:nonidealnfw}

After characterising the nearly pure spherical NFW model, we test the accuracy of the empirical expansion on two non-ideal NFW test cases. In the first, we test the recovery of forces in an idealised cored halo particle distribution. The central profile of dark matter halos is still a matter of considerable debate, both in terms of structure formation as well as evolutionary processes that may form cored profiles from cusps. Our procedure to test the ability of our BFE method to represent such an evolution from a cusp profile to a cored profile is as follows. We first realise a new halo with the same parameters as in Section~\ref{subsubsec:basenfw}, except with $r_c=0.006R_{\rm vir}$. We then expand the forces using the basis derived in Section~\ref{subsubsec:basenfw} using the $r_c=0.0002R_{\rm vir}$ basis. In this case, the forces remain analytic, so we may again compute \(\mu_{\Delta_r, \Delta_\theta, \Delta_\phi}\) and \(\sigma_{\Delta_r, \Delta_\theta, \Delta_\phi}\). We find, as expected, that the \((l,n)=(0,1)\) function is no longer a near-perfect representation of the force. Given the structure of the test, we know that the monopole terms should be able to reproduce the cored distribution. 

Unlike in the idealised case, the accuracy of the forces first increases for \(l=0,n>1\), with a minimum in the overall force error at \(l=0,n=10\), roughly in agreement with the node spacing required to resolve the core based on visual inspection of the functions. Using the principal results of Table~\ref{tab:forceerror} as a comparison, we find that the overall force errors remain largely unchanged for \(l=6,n=24\): \(\mu_{\Delta_r}=6.5\times10^{-2}\%\), \(\mu_{\Delta_r}(r<0.01R_{\rm vir})=7.8\times10^{-2}\%\). As expected for a spherical distribution, the force error in the angular coordinates remains unchanged. We take this as evidence that the evolution of purely spherical core formation would be well-resolved using {\sc exp}.

The asphericity of dark matter halos is also still a matter of debate, although it is reasonable to assume that dark matter halos in nature are not perfectly spherical. As a second example of a non-ideal model, we test the force accuracy in an oblate spherical distribution. In practice, we use the base model from Section~\ref{subsubsec:basenfw} and make the substitution \(r\to\sqrt{R^2 + q^2z^2}\). When \(q=1\), the halo is spherical. Making \(q\) smaller results in successively more oblate halos. In the general case of aspherical halos, we can no longer use analytic forces. We, therefore, turn to a high-order multipole expansion. The analytic forces are computed by a high-precision multipole expansion and subsequent analytic differentiation by expanding the disc density with a high number of radial and elevation integration knots. In equations~(\ref{eq:raderr_sph})--(\ref{eq:phierr_sph}) the `analytic' forces are now computed from the multipole expansion. We test three oblate halos: \(q={0.95,0.9,0.85}\).

Again comparing to the principal results of Table~\ref{tab:forceerror}, we find that the overall force errors only increase modestly, even for the most extreme \(q=0.85\) case computed at \(l=6,n=24\): \(\mu_{\Delta_r}=8.99\times10^{-2}\%\), \(\mu_{\Delta_r}(r<0.01R_{\rm vir})=9.5\times10^{-2}\%\). In this test, the true polar force error is now non-zero, and we find a corresponding modest increase in the force error: \(\mu_{\Delta_\theta}=2.1\times10^{-3}\%\), \(\mu_{\Delta_\theta}(r<0.01R_{\rm vir})=4.6\times10^{-2}\%\). The true azimuthal force is still zero, and we find that the errors are unchanged.

As an example of the characteristic accuracy and root variance curves, Figure~\ref{fig:sph_accuracy_compare} shows \(\mu_{\Delta_r}\) (panel b) and \(\sigma_{\Delta_r}\) (panel c) as a function of radius for the three different models discussed: near-ideal NFW (the subject of section~\ref{subsubsec:basenfw}), the cored NFW, and the oblate NFW. The realised (ideal) spherical-average densities for each model are shown as solid (dashed) curves in panel a. We compare the {\sc EXP} (solid curves in panels b and c) with the {\sc gadget-2} forces (dashed curves in panels b and c). In general, we find that {\sc exp} exhibits smaller force errors (mean force accuracy, upper panels), and significantly smaller root variance (force accuracy variance, lower panels), when compared to {\sc gadget-2} forces computed for the same particle distribution.

While Figure~\ref{fig:sph_accuracy_compare} draws considerable attention to the centre of the galaxy -- which is important -- we stress that the majority of the particles are in the outer, high-accuracy regions of {\sc exp}-calculated forces. Further, the inner regions of the realised particle distribution are not smooth (panel a of Figure~\ref{fig:sph_accuracy_compare}, motivating the use of alternate sampling techniques, see Section~\ref{subsubsec:multimass}), such that the force inaccuracy is dominated by discreteness in the particle realisation. The small exception to the high-accuracy forces in the outer halo is the force accuracy at \(\log(r)>-1.\) For the oblate halo, owing to the mismatch of the outer profiles, the {\sc exp}-calculated forces are unable to achieve the same force accuracy as for the other cases, where the outer profiles match the basis.

In both the cored and oblate halo tests, we find that the variance curves remain largely unchanged, and as such the noise in the BFE force realisation is significantly smaller than that in a tree-gravity realisation (an order of magnitude in the halo, on average). While it is beyond the scope of this paper, a future comparison of differences in evolution resulting from aphysical noise would be a fruitful project.

\subsubsection{Multimass halos and accuracy} \label{subsubsec:multimass}

Interesting dynamics often takes place at the centre of the dark matter halo. Unfortunately, owing to finite sampling of the halo, the centre is often under-sampled. One scheme to improve the sampling in the inner halo is the {\sl multimass} scheme: the number density of dark matter particles may be biased towards smaller radii by adjusting the per particle mass.

As described in \citet{petersen21}, one creates two distribution functions as a function of energy \(E\): one that corresponds to the true desired mass (or density) distribution function (the mass distribution function, \(f_{\rm mass}\)) and one that corresponds to the desired number density (the number distribution function, \(f_{\rm number}\)). Particles are realised from \(f_{\rm number}\) with uniform mass \(m_{\rm number}\) with the mass for each particle rescaled such that \(m_{\rm mass} = m_{\rm number}f_{\rm mass}(E)/f_{\rm number}(E)\). The particles then match the mass distribution of the true desired distribution function with the number density of \(f_{\rm number}\). A typical choice for the target number density profile is a simple power-law distribution, $n_{\rm halo}\propto r^{-\alpha}$, where $\alpha\in[2,3]$. For example, in \citet{petersen21}, we chose \(\alpha=2.5\), resulting in a factor of 100 increase in particles in the inner halo \(r<0.05R_{\rm vir}\). 

However, the multimass scheme changes the bias in the estimate of the coefficients for the lowest-order function. The \(l=n=0\) spherical basis function derived using the Sturm-Louiville method only matches the potential estimate exactly for the equal mass particle case, where the measure used to derive the basis, \(d^3 x  \rho\), is proportional to the particle mass. Empirically, the coefficient bias varies with the number distribution for a fixed mass distribution. Thus, we are left with an optimisation problem that balances the desire for force accuracy against the desire for a finely-sampled phase space in dynamical regions of interest. Selecting the optimal multimass weighting depends on the problem at hand. One has to be careful about the bias and some experimentation and bias calibration needs to be part of any multimass simulation strategy. For example, we find that for the configuration used in \citet{petersen21}, the mean force error for the \((l=0,n=1)\) term increases to \(\mu_{\Delta_r}=6.0\times10^{-2}\%\), but the addition of extra terms somewhat mitigates this bias. For \(l=6,n=24\), we find \(\mu_{\Delta_r}=7.6\times10^{-2}\%\).

\subsection{Cylindrical Expansions}\label{subsec:cylindrical}

While the lowest-order function in an initially spherical model is a near-exact match to the density and potential, the truncation of the series and conversion to meridional space may result in deviations from the true density and potential functions for cylindrical expansions. We validate the cylindrical basis by comparing the {\it expanded} forces to {\it analytic} forces. As above for the non-spherical cases, the analytic forces are computed by a high-precision multipole expansion and subsequent analytic differentiation by expanding the disc density using a high number of radial and elevation integration knots. We then compute the relative difference for the forces at the position of each particle: 
\begin{eqnarray}
\Delta_R(R,z) &= \frac{\left|F_{R,~{\rm expanded}}(R,z) -F_{R,~{\rm analytic}}(R,z)\right|}{F_{R,~{\rm analytic}}(R,z)} \label{eq:raderr}\\ 
\Delta_z(R,z) &= \frac{\left|F_{z,~{\rm expanded}}(R,z) -F_{z,~{\rm analytic}}(R,z) \right|}{F_{z,~{\rm analytic}}(R,z)} \label{eq:verterr}\\
\Delta_\phi(R,z) &= \frac{\left|F_{\phi,~{\rm expanded}}(R,z) -F_{\phi,~{\rm analytic}}(R,z) \right|}{\sqrt{F_{R,~{\rm analytic}}^2+F_{z,~{\rm analytic}}^2+F_{\phi,~{\rm analytic}}^2}(R,z)}. \label{eq:phierr} \end{eqnarray}
where $F_R$ is the cylindrical radial force, $F_z$ is the vertical force, $F_\phi$ is the azimuthal force, and the subscripts `expanded' and `analytic' refer to the BFE forces and the multipole expansion forces, respectively. We normalise the azimuthal force by the total force to avoid zeros in the azimuthal force that result from symmetries in the test density configurations. We test two configurations: a pure exponential disc and a model designed to approximate the density distribution of an evolved disc that includes a bar. In both tests, we use the same particle realisation to compute the forces so that the results may be compared independent of realisation noise.

As in the spherical case, we compare with the tree-code gravity of {\sc gadget-2} \citep{springel05}. We additionally compute the relative force errors for direct-summation gravity. For the direct-summation gravity, we use a ring-algorithm \citep{makino02} implemented in the {\sc exp} framework. We again choose a cubic spline kernel\footnote{Using the Plummer softening kernel has little impact on $F_R$ or $F_\phi$, but changes the distribution of the $F_z$ errors appreciably.} with a softening length of $h=0.000952R_{\rm vir}=0.0952R_d$. The relative forces are computed using equations~(\ref{eq:raderr})--(\ref{eq:phierr}), replacing the expanded forces with the direct-summation and tree-gravity forces. As discussed above, for {\sc gadget-2}, we input the softening length $\epsilon=0.00034R_{\rm vir}=0.034R_d$.

\subsubsection{Exponential Disc Test}\label{subsec:exponentialdisc}

\begin{figure*} \centering \includegraphics[width=6.5in]{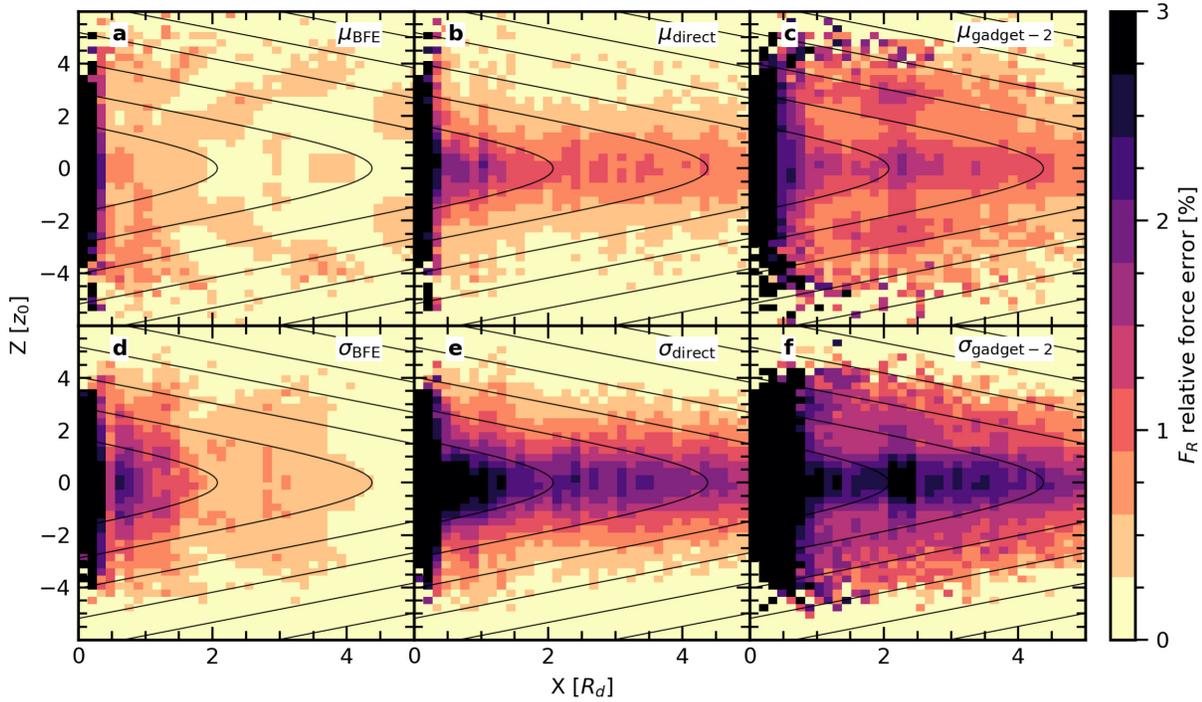} \caption{\label{fig:icRerror} Median (panels a,b,c) and root variance (panels d,e,f) of the absolute value of the relative radial force errors in the meridional plane for an $N=10^6$ exponential disc model using three gravity solvers, as labelled. Black curves show contours of constant density.} \end{figure*}

\begin{figure*} \centering \includegraphics[width=6.5in]{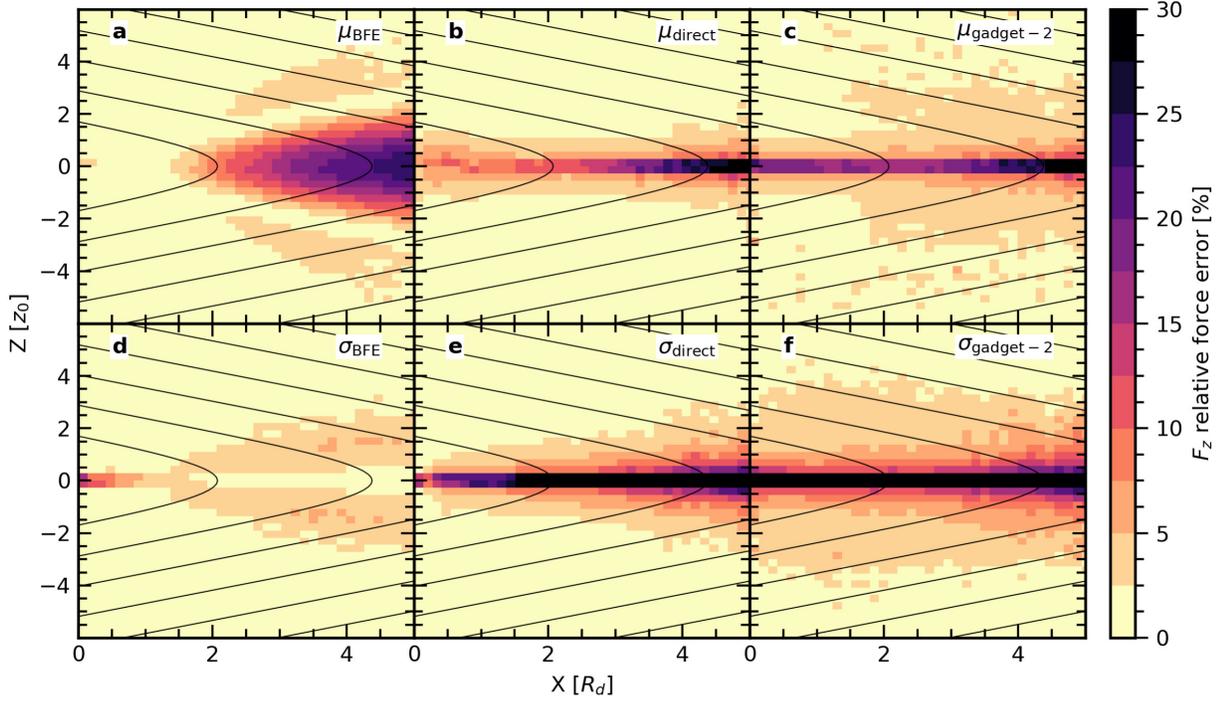} \caption{\label{fig:icZerror} The same as Figure \ref{fig:icRerror} but showing the relative $z$ force errors.} \end{figure*}

\begin{figure*} \centering \includegraphics[width=6.5in]{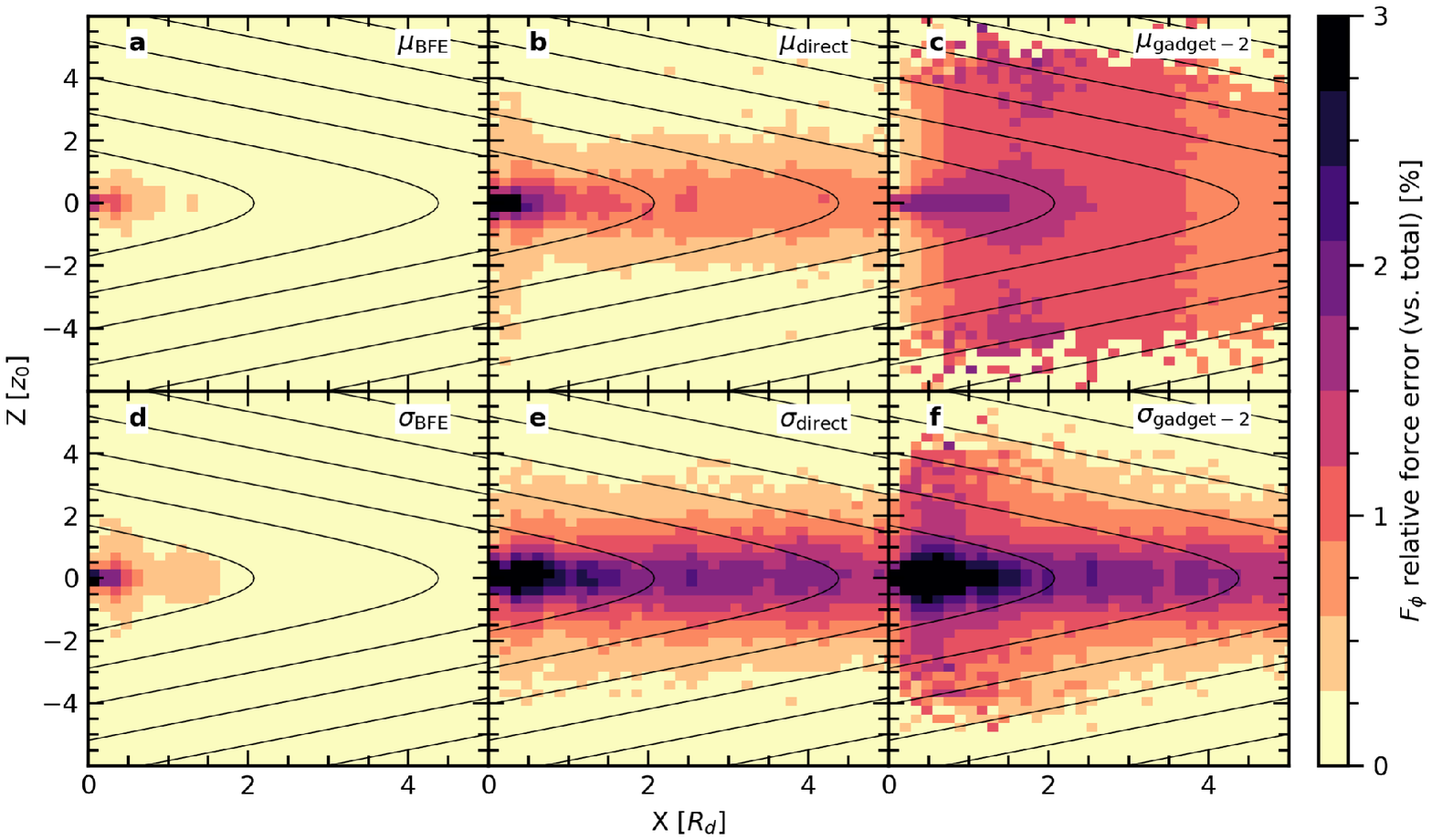} \caption{\label{fig:icPerror} The same as Figure \ref{fig:icRerror} but showing the relative $\phi$ force errors. The force errors are now relative to the total forces.} \end{figure*}

Our exponential disc is parameterised as an exponential in radius and an isothermal ${\rm sech}^2$ distribution in the vertical dimension: \begin{equation} \rho_d(r,z) = \frac{M_{\rm d}}{8\pi z_0R_d^2} e^{-r/R_d} {\rm sech}^2 (z/z_0) \label{eq:exponentialdisc} \end{equation} where $M_d$ is the disc mass, $R_d$ is the disc scale length, and $z_0$ is the disc scale height, which is constant across the disc. For this experiment, we test a $z_0/R_d = 1/10$ scale height to scale length ratio. We first test the same exponential disc model used to condition the basis. 

In Figure~\ref{fig:icRerror}, we show the median and variance of the absolute value radial force error for binned particles using the three gravity solvers. We have scaled the force errors by a 100 to show the results as percentages.  We take advantage of the axisymmetry of the disc and bin the relative force errors in the meridional plane. The left column shows the results for the BFE, using the parameters taken from the simulation presented in \citet{petersen21}: \(m_{\rm max}=6,~n_{\rm max}=12\). The centre column (panels b and e) shows the results for the direct summation solver, and the right column (panels c and f) shows the results for {\sc gadget-2}.

For the BFE, we find that $\Delta_R(R,z)<4$\% everywhere, with the largest errors at $R<0.2R_d$ for the basis function expansion with $n_{\rm max}=12$.  However, the places with the largest radial force errors do not contain many particles, and the median force errors for all particles in the initial disc distribution are $\Delta_R=0.52$\%. For both direct-summation and {\sc gadget-2}, the largest force errors occur near the disk plane, which has the highest particle density. The median radial force errors are 1.02\% and 1.50\% for direct summation and {\sc gadget-2}, respectively. Additionally, the errors from the quadrupole-order multipole used to compute long-range forces in {\sc gadget-2} are apparent in panel c, when compared to panel b.
 
The vertical force errors in {\sc exp}, shown in panel a of Figure~\ref{fig:icZerror} are modestly worse than the radial force errors, but in expected ways: the force accuracy declines in the plane at larger radii, where the vertical forces are small.  Once again, the regions with the largest force errors are not populated by many particles, and the median $z$ force error $\Delta_z=2.2$\%. Direct summation generally has larger radial and $z$ force errors than {\sc exp} within a couple of disk scale lengths but {\sc exp} has larger $z$ force errors at larger radii. The corresponding median vertical force errors for direct summation and {\sc gadget-2} are 5.1\% and 6\%, respectively. 

In an axisymmetric system, the BFE excels at minimising spurious forces in azimuth, as seen in panel a of Figure~\ref{fig:icPerror}. Owing to the softening kernel and local particle fluctuations, both direct summation and {\sc gadget-2} show large non-zero azimuthal forces. When normalised by the total force, the fractional azimuthal force is 0.44\%, 1.02\%, and 4.4\% for BFE, direct summation, and {\sc gadget-2}, respectively. The particularly poor {\sc gadget-2} forces appear to be a consequence of the tree construction, as the pattern in the force errors corresponds with the pattern in the radial force errors.

The variance at each point in the meriodional plane, shown in the lower panels of Figures~\ref{fig:icRerror}-\ref{fig:icPerror}, demonstrates the minimal and stable noise properties of {\sc exp}. One can see that the regions of largest force error in {\sc exp} have very small variance while for the direct-summation and tree gravity solvers the places with the largest force errors also have the largest variance. These are also the places that are most populated by disk particles. The basis function expansion, {\sc exp}, has a stable pattern that will introduce a constant bias rather than a changing noise pattern, as in direct-summation or tree gravity, i.e. since the systematic force errors from the truncated series of basis functions are time-independent, they have a minor effect on secular evolution. The BFE will still be affected by fluctuations, but the systematic bias from the particular basis truncation has a time-independent component.

We have verified that the truncation errors in $\Delta_{\{R,z,\phi\}}(R,z)$ tend to zero as $n_{\rm max}$ increases. One may improve the vertical force resolution specifically by constructing a basis with higher $l_{\rm initial}$, which increases the angular resolution and may create functions that more closely resemble the outer disc. However, as the vertical force is small at larger radii, we do not believe this affects typical simulations \citep[e.g.][]{petersen19,petersen21}. We have tested the evolution using bases realised with higher $l_{\rm initial}$ and find no appreciable differences.

\subsubsection{Bar-and-disc model test}\label{subsec:model2bar}

\begin{figure*} \centering \includegraphics[width=6.2in]{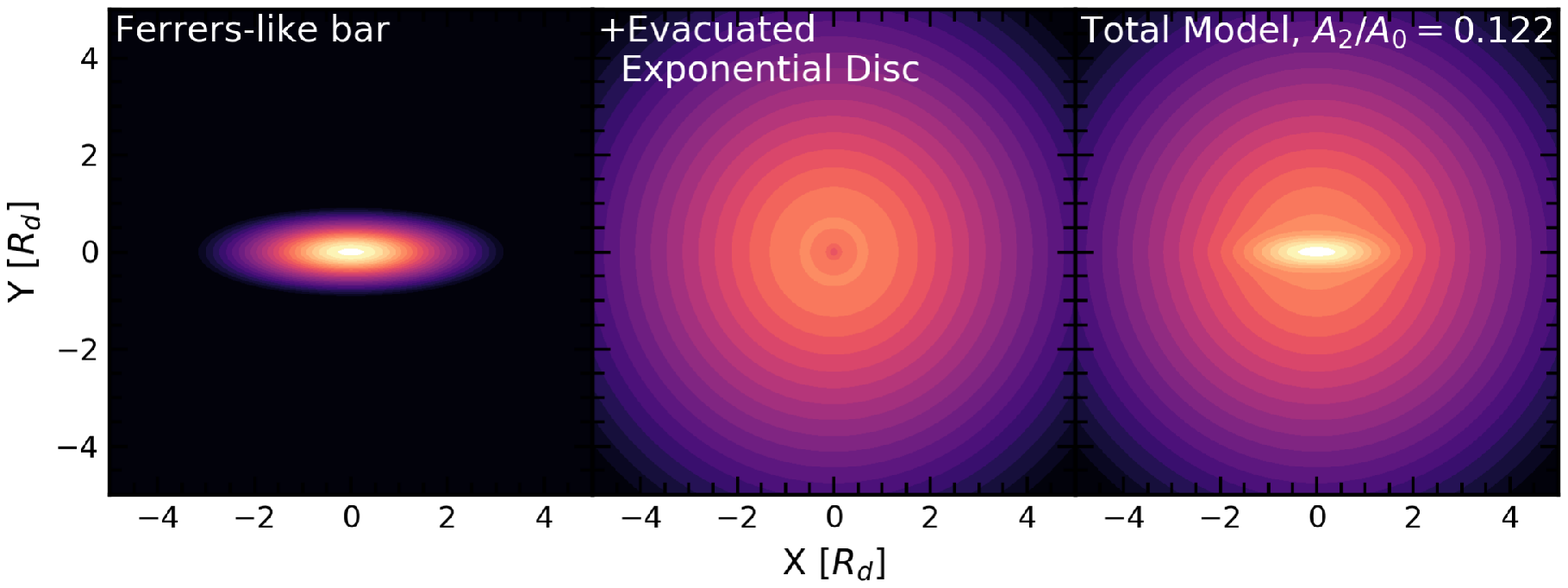} \caption{\label{fig:model2} Model for the bar-and-disc system, where the panels show the bar model, the evacuated disc model, and the total model (from left to right).} \end{figure*}

\begin{figure*} \centering \includegraphics[width=6.5in]{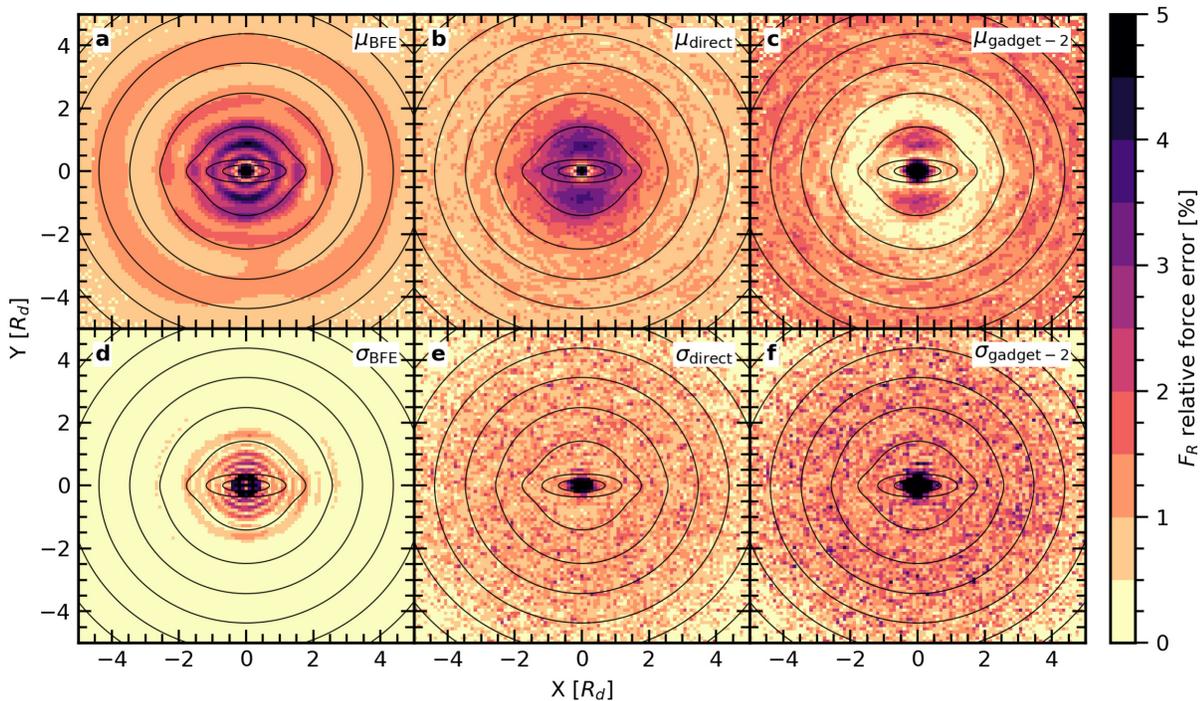} \caption{\label{fig:model2Rerror} Mean (panels a,b,c) and root variance (panels d,e,f) of the absolute the value of the relative radial force errors for an $N=10^6$ bar-and-disc model using the three gravity solvers as labelled, in the $x-y$ plane, for particles $|z|<0.5z_0$. Black curves show contours of constant density.} \end{figure*}

\begin{figure*} \centering \includegraphics[width=6.5in]{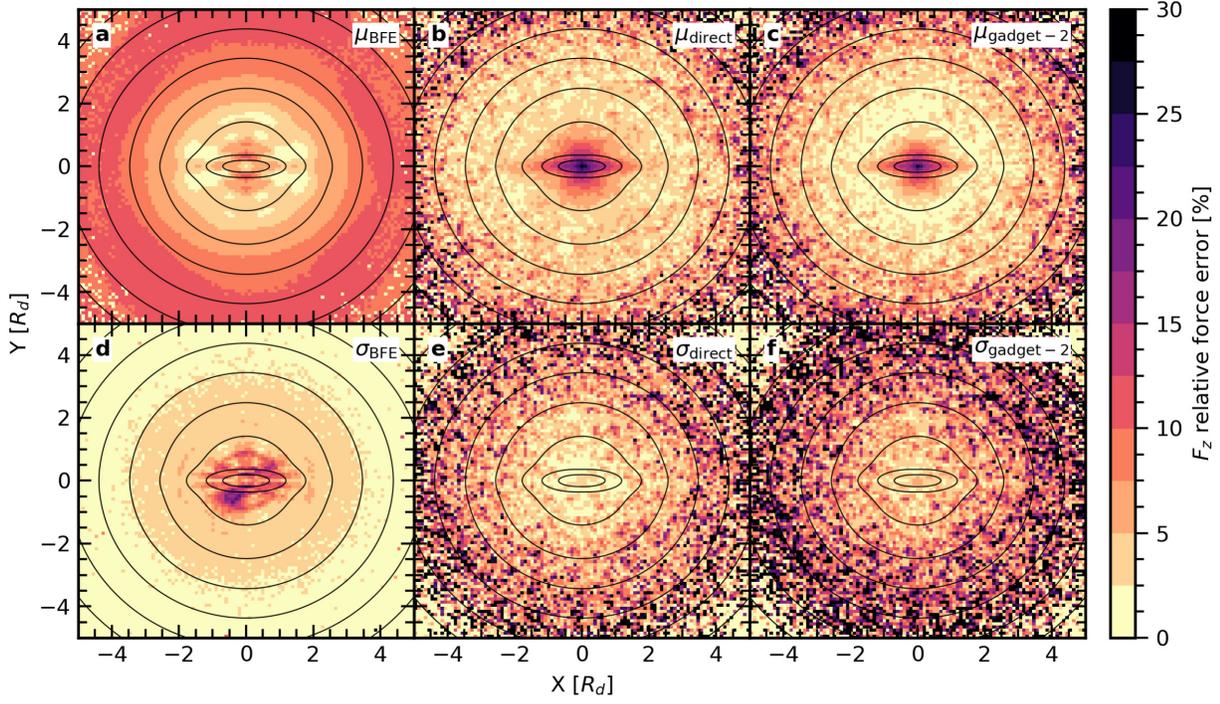} \caption{\label{fig:model2Zerror} The same as Figure \ref{fig:model2Rerror} but showing the relative $z$ force errors for particles $0.5z_0<z<1.5z_0$} \end{figure*}

\begin{figure*} \centering \includegraphics[width=6.5in]{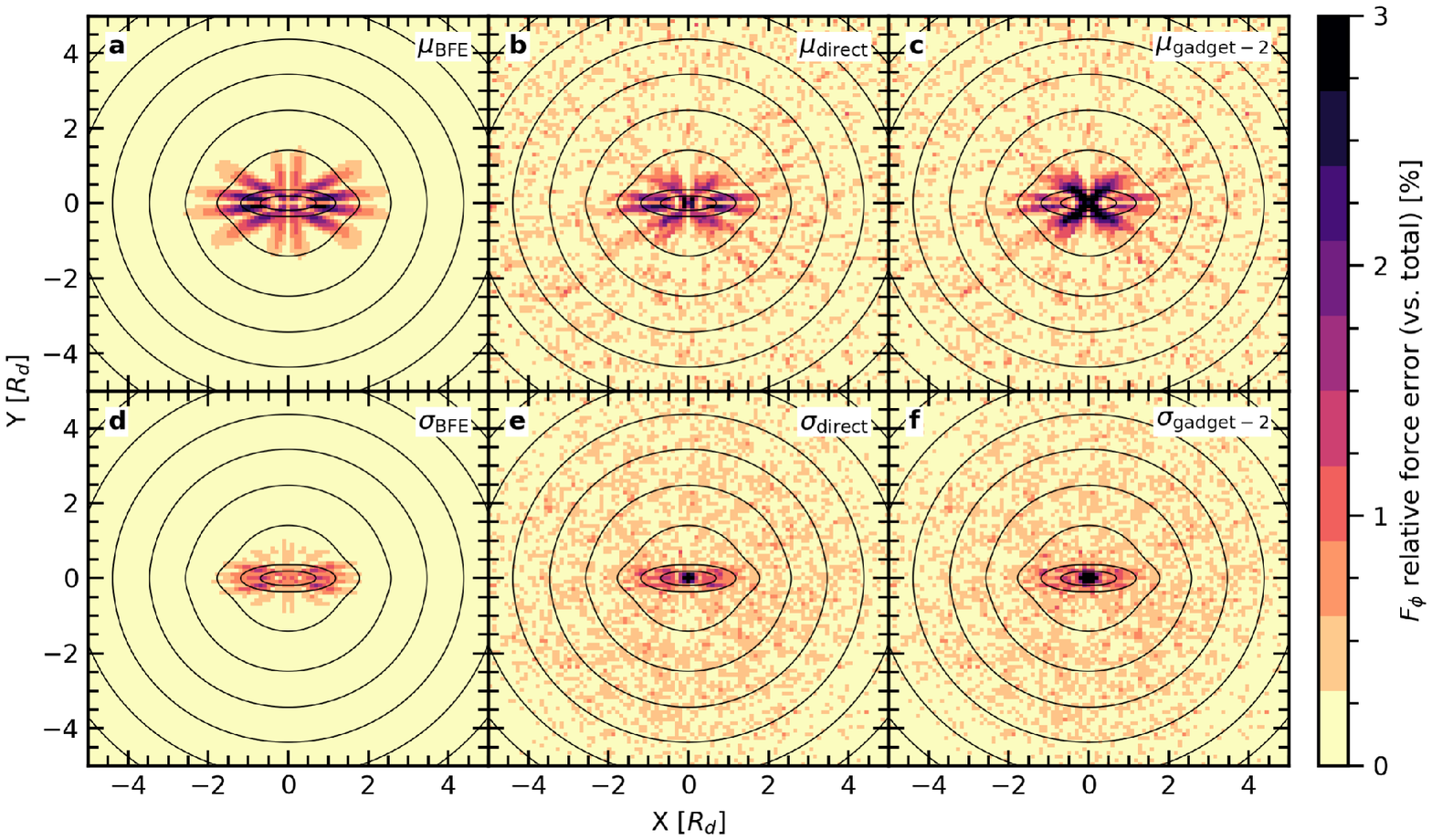} \caption{\label{fig:model2Perror} The same as Figure \ref{fig:model2Rerror} but showing the relative $\phi$ force errors. The force errors are now relative to the total forces.} \end{figure*}

Discs often develop strong non-axisymmetric features such as spiral arms and bars.  To test forces in a non-axisymmetric disc, we design a model that resembles the late-time configuration of the cusp simulation from \citet{petersen21} to test the basis forces. We show the model in Figure~\ref{fig:model2}. The model consists of two components: a Ferrers-like bar and an evacuated exponential disc. 

We consider a bar profile with the Ferrers ellipsoid form given by a generalised formula that is a softened power law in elliptical coordinate $m$,
\begin{equation}
    \rho_{\rm bar~model} = \rho_c(1+m^\mu)^\nu
\label{eq:softferrers}
\end{equation}
where we define $m$ as above and we choose $\mu=2$ and $\nu=-4$. Although the density in equation~(\ref{eq:softferrers}) formally has infinite extent, the density is steep and the profile quickly converges to its asymptotic value. We use the same axes as in the Ferrers bar test above. While this model does not result in analytic forces, it is a better numerical test of the force accuracy than a traditional Ferrers bar. We show the density of the bar itself in the left-most panel of Figure~\ref{fig:model2}.

We also require an exponential disc model to complete the test density model. We modify the axisymmetric disc to remove the particles that are now a part of the bar. To do so, we parameterise the disc with the initial exponential disc and then `evacuate' the central region using an inverted exponential disc (the negative of eq. \ref{eq:exponentialdisc}) with the same mass as the bar and scale height as the initial disc. We tie the scale length of the disc to the bar and choose $R_{d,{\rm inverted}}=a/3$, a value which results in a relatively small radial mass rearrangement when combined with the bar model. We show the evacuated exponential disc in the middle panel of Figure~\ref{fig:model2}, and the total model in the right panel of Figure~\ref{fig:model2}. 

We generate a realisation for the disc and bar model through rejection sampling. Using the projected surface density of the realised particles, we find the Fourier-measured strength of the bar, $A_2/A_0=0.122$, which is typical for the bars found in \citet{petersen21}.

For each particle in the bar-and-disc model, we compute the relative force accuracy (eqns.~\ref{eq:raderr}-\ref{eq:phierr}) for the BFE-computed forces using the same basis as for the disc model, the direct-summation forces, and {\sc gadget-2}-computed forces. In Figures~\ref{fig:model2Rerror}-\ref{fig:model2Perror}, we show the force errors as in Figures~\ref{fig:icRerror}-\ref{fig:icPerror}. Owing to the non-axisymmetric structure of the bar-and-disc model, we now bin the relative force errors in $x-y$ space, selecting particles in a vertical slice. For the radial and azimuthal forces, we select all particles that satisfy $|z|<0.5z_0$. For the vertical forces, we select all particles that satisfy $0.5z_0<z<1.5z_0$.

The results are much the same as in the exponential disc case, with modestly increased errors. The median radial force errors for the three potential solvers are 1.4\%, 1.9\%, and 3.0\% for the BFE, direct summation, and {\sc gadget-2} cases. The median vertical errors are 4.5\%, 5.5\%, and 7.4\%, and the median azimuthal errors are 1.0\%, 1.8\%, and 1.9\% for the solvers (in the same order). The BFE force approximation results in a distinct spatially-coherent low-level bias pattern resulting from the basis truncation, but has relatively little variance. The direct-summation and {\sc gadget-2} forces feature both bias and significant variance resulting from the softening. 

While the bias pattern is most apparent by eye in the BFE median panels (panel a), an edge will produce features in any gravity solver: in BFE, it is ringing by truncation (analogous to the Gibbs phenomenon), in direct and tree-based gravity, is is bias from oversmoothing. Thus, while the BFE shows a clear, low-level ringing as a result of the bar edges, direct summation and tree-based gravity both show error features related to the bar geometry. The effect is best understood as a kernel mismatch between the basis functions in the case of BFE, and the smoothing kernel in the case of direct summation and {\sc gadget-2}.

\begin{figure*} \centering \includegraphics[width=6.2in]{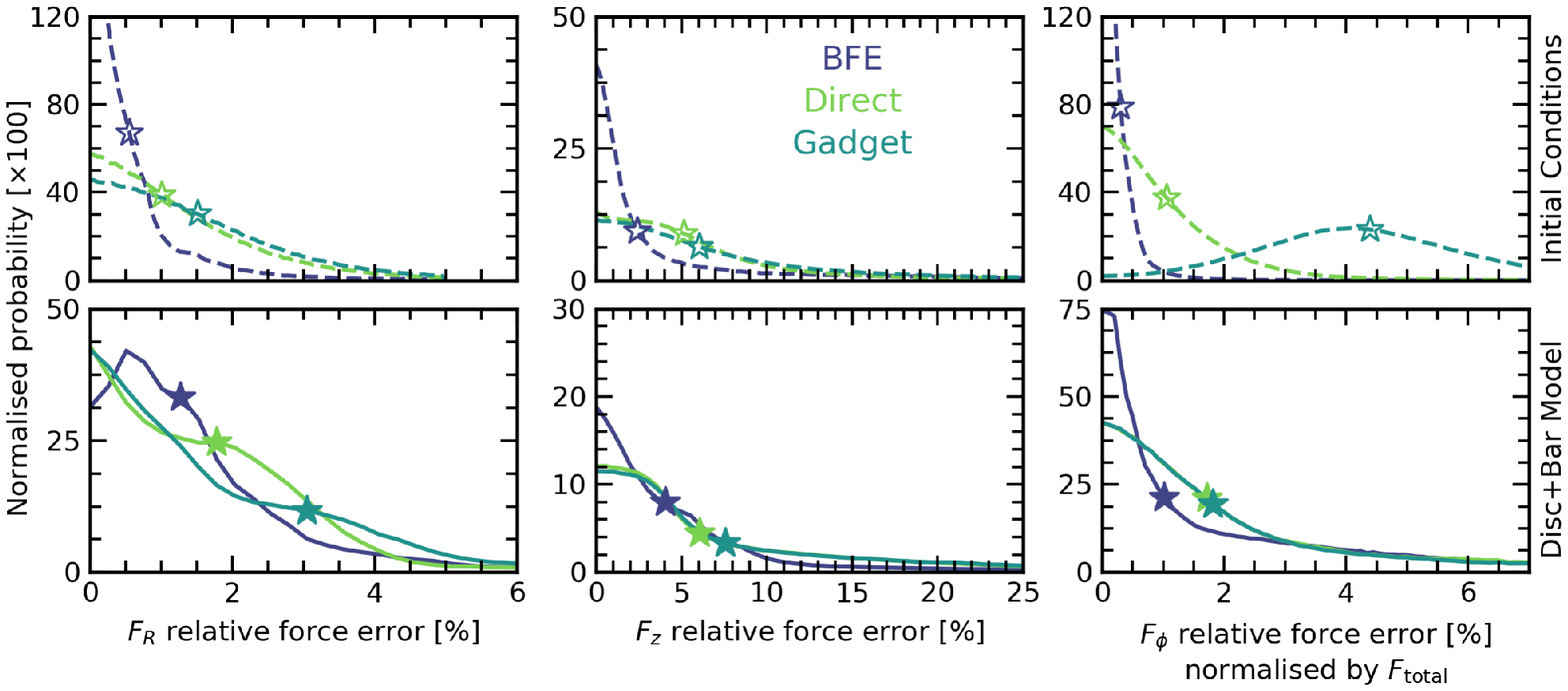} \caption{\label{fig:multipoleforcehistograms} Normalised probability distributions of the relative force errors for an exponential disc particle distribution (panels a-c) and the bar+disc model (panels d-f), computed using the three different gravitational force evaluation techniques as indicated. Stars indicate the median value for the total particle distribution.} \end{figure*}

We summarise the results of these force error tests in Figure~\ref{fig:multipoleforcehistograms}, where we show the entire normalised probability distribution of force errors in each dimension, for both the initial disc (upper row) and bar-and-disc model (lower row). We summarise the overall force error by marking the median with a star. For both models studied here, an exponential disc with and without a bar, BFE has a lower median relative force error when compared to direct-summation or {\sc gadget-2}. The reasons are straightforward; the BFE is able to accurately determine forces in high density regions with little variance. The spatial regions with the largest bias are those with the fewest particles, i.e. where the basis has the least support, which in turn affects a relatively small number of particles. In contrast both the direct and the tree method have their largest errors and force variance in regions that contain many particles.

\subsection{Coefficient significance}\label{subsec:coefficientsignificance}

The high-order basis functions in {\sc exp} contain information about small spatial scales and require a large number of particles to compute accurate coefficients\footnote{Recall that the coefficient amplitude for a particular eigenfunction $u_\mu$, corresponding to the potential function $\phi_\mu$, may be written as $a_\mu = \int d\mathbf{x}\, \rho(\mathbf{x}) \phi_\mu(\mathbf{x})$ (cf. equation~\ref{eq:amplitude}).}.  For a fixed number of particles \(N\), there will be some maximum order beyond which the coefficients are noise dominated (low coefficient significance).  We now briefly outline our procedure for determining coefficient significance.

The particle distribution itself correlates the error in the various coefficients, so an analysis of a coefficient covariance matrix  \(\mbox{cov}(\mathbf{a})\) is necessary to determine independent degrees of freedom.  The covariance of the potential basis functions \(\phi_\mu\) and \(\phi_\nu\) is given by \begin{equation} \mbox{cov}(\mathbf{a})_{\mu\nu} = \int d\mathbf{x} \rho(\mathbf{x}) \phi_\mu(\mathbf{x}) \phi_\nu(\mathbf{x}) - a_\mu a_\nu. \label{eq:cmuvarint} \end{equation} 
In {\sc exp}, we compute the signal-to-noise ratio, \(S\), for each independent degree of freedom using a bootstrap resampling technique as follows.  We partition the \(N\) particles into \(J=\sqrt{N}\) partitions with \(J\) particles each, and compute the coefficients for each basis function in each partition, \(\hat{a}_{\mu,j}\), where \(j=1,\ldots,J\). We then construct a covariance matrix for each azimuthal order for the set of \(J\) coefficients.  We use a singular value decomposition to compute the eigenfunctions and eigenvalues of this covariance matrix.  Transforming the original coefficients to the new basis implied by the eigenfunctions yields a new set of \emph{rotated} coefficients \(\tilde{a}_{\mu}\) (where we use \(\tilde{\cdot}\) to denote quantities in the rotated space). The eigenvalues of the covariance matrix, \(\tilde{b}_{\mu}\), are an estimate of the variance in the projected coefficients uncorrelated by the original basis.  We then define \(S_\mu = \tilde{a}_{\mu}/\sqrt{\tilde{b}_{\mu}}/J\).

In practice, the covariance matrix is diagonally dominated. This implies that the eigenfunctions of our covariance matrices smooth the original coefficients about the diagonal.  Thus, one may also make a fair estimate of the signal-to-noise ratio in the \emph{unrotated} coefficient $\hat{a}_\mu$ using \(S_\mu\).  Empirical tests using this bootstrap resampling technique on the simulation outputs of \citet{petersen21} suggest that the noise floor in both a spherical (halo) and cylindrical (disc) component are reached at approximately \(S_\mu=4\).  Specifically, we estimate the signal-to-noise floor by examining the run of \(\tilde{a}_{\mu,j}\) for high \(j\) and recording the value of \(S\) where \(\tilde{a}_{\mu,}\) is no longer coherent in time, but fluctuates in a white noise fashion. In the simulations presented in \citet{petersen21}, at most the highest two (four) radial orders fall below the estimated noise floor in the cylindrical (spherical) basis for long stretches of time in their simulations. Additionally, in the spherical basis, some higher-order angular terms exhibit low significance. \citet{petersen21} did not truncate their basis owing to the small number of obviously low-significance terms.

\begin{figure*} \centering \includegraphics[width=5.7in]{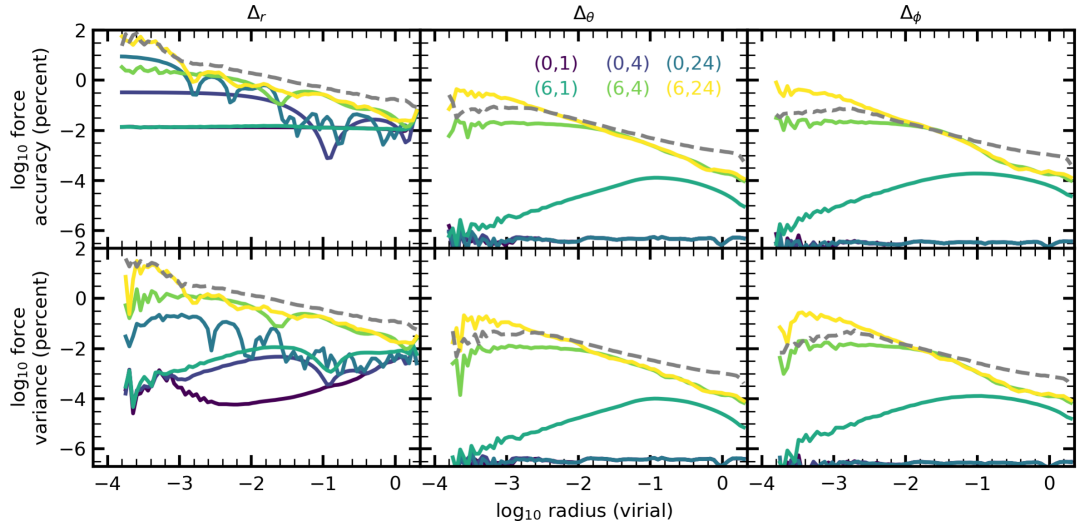} \caption{\label{fig:sph_accuracy} Measurement of the logarithm of mean force accuracy (upper panels) and logarithm of the root variance of force accuracy (lower panels) as a function of the logarithm of the radius for various \((l,n)\) combinations (coded by colour). From left to right, the columns are radial, polar, and azimuthal forces. We show the forces computed using {\sc gadget-2} as a grey dashed line for reference.} \end{figure*}

However, the generic presence of low-significance terms naturally suggests that implementing a `coefficient smoothing' algorithm would reduce both the bias and variance in the force computation by reducing or eliminating the contribution from noise dominated coefficients \citep{weinberg96}.  As an example, consider the idealised spherical NFW test from Section~\ref{subsubsec:basenfw}. Figure~\ref{fig:sph_accuracy} shows \(\mu_{\{\Delta_r, \Delta_\theta, \Delta_\phi\}}\) and \(\sigma_{\{\Delta_r, \Delta_\theta, \Delta_\phi\}}\) as a function of radius for different combinations of \(l_{\rm max},n_{\rm max}\) to show the changing force accuracy as more terms are included. As the basis is designed to resemble the equilibrium profile, the lowest-order term is a high-precision match the forces. When further terms are added, the force accuracy slowly becomes worse and the variance slowly increases. In the case of this equilibrium model, where the lowest-order basis term nearly perfectly describes the system, it is evident that the optimal smoothing would reduce the contribution to the forces from the other terms. However, the terms are needed to resolve later evolution in the system. For the present application, we find that even with the addition of noise from low-significance coefficients, the forces are still (on average) significantly more accurate, with less force variance, than the comparable {\sc gadget-2} computation. 

As a test, we perform the method above to rotate the coefficients and compute the signal-to-noise for the NFW model in Section~\ref{subsubsec:basenfw}. We use a simple trimming procedure (eliminating all coefficients below a specified signal-to-noise threshold) and scan through signal-to-noise cuts from 1 to 10. We find a minimum at a signal-to-noise of 5, where we improve the overall radial force accuracy in the model from 6.4e-2\% to 4.5e-2\%, with comparable improvements at all radii. For the angular components of the forces, we find force accuracy improvements of greater than an order of magnitude: the azimuthal force accuracy improves from 4.9e-4\% to 4.4e-5\%, and the polar forces improves from 3.5e-4\% accuracy to 2.2e-5\% accuracy. One may reasonably expect with a more advanced smoothing algorithm in place, the typical accuracy and variance may be improved by an order of magnitude or more.

\subsection{Force accuracy summary} \label{subsec:forcesummary}

The force accuracy tests presented here illustrate the features of the approximations inherent in BFE, direct-summation, and tree gravity $N$-body techniques. In particular, we learned:
\begin{enumerate}
\item BFE provides an accurate representation of the potential and force for distributions that evolve only moderately from their initial conditions.   In particular, BFE should perform well for the quiescent evolution of disc galaxies, including core formation or the formation of a strong bar.
\item For galaxy disks, including those with strong bars, BFE has the lowest median relative force errors followed by direct summation. Tree-based gravity ({\sc gadget-2}) performs modestly worse than direct summation. In the dark matter halo, BFE has consistently smaller median relative force errors.
\item Direct-summation and tree ({\sc gadget-2}) gravity is less accurate in higher-density regions owing to gravitational softening. BFE force accuracy is independent of the local density except in low-density regions where the potential is poorly constrained.
\item Sharp changes in density are difficult for all the gravity solvers to represent. In the case of BFE, this results in bias patterns that resemble the underlying functions; in the case of direct-summation and tree gravity, a sharp density change results in oversmoothing and the edge being poorly determined.
\item To evaluate force accuracy, one cannot compare a BFE to direct-summation forces owing to the different biases and variance patterns in the solvers. The only fair force comparison is to compare both BFE and direct-summation forces to a true force.  Here we used either analytic expressions and/or a high-order multipole expansions to compute accurate forces.
\end{enumerate}

Our results immediately prompt the question of whether modestly biased, but low noise forces are better for resolving secular evolution as compared to small bias, higher noise forces. Unfortunately, a full study of the bias-variance tradeoff in different potential solvers is beyond the scope of this work, but remains an open question in high-resolution dynamics.

\section{Evolution} \label{sec:expevolution}

With the computed basis or bases in hand, we may then proceed to evolve the dynamical system in time. {\sc exp} uses a symplectic leapfrog integrator following a `kick-drift-kick' scheme \citep{quinn97}. One can show through direct computation of the Taylor series expansion of the Hamiltonian that the phase-space accuracy in leapfrog integration after a single timestep is $\mathcal{O}(h^3)$ \citep{yoshida90}. For computational purposes, the leapfrog integrator is an inexpensive integrator that requires only one evaluation of the potential per timestep, and no storage of previous timesteps, making the algorithm computationally economical. Leapfrog is also time reversible. Time-reversibility is a constraint on the phase-space flow that, like symplecticity, suppresses numerical dissipation, since dissipation is not a time-reversible phenomenon\footnote{See the discussion in \protect{\citet{springel05}} regarding individual particle timesteps and symplecticity, but see also \protect{\citet{hernandez19}} for a cautionary note in specific cases.}.

\subsection{Timesteps}

The choice of timesteps in an $N$-body simulation has been discussed extensively in the literature \citep[see, e.g.,][]{dehnen17}, with the principal goal of avoiding artificial energy dissipation and conserving angular momentum. To this end, we have extensively tested the timesteps in {\sc exp} and developed criteria that meet the required conservation precision.

{\sc exp} requires an input master timestep, which is the largest timestep a particle may be evolved for at a given step. In general, we choose timesteps so there are at least 100 steps over an orbital period \citep{weinberg07a}.  Specifically, we compute three time scales for each particle, at each timestep, choosing the timestep that gives the most stringent value: \begin{enumerate} \item The force time scale: \(|\mathbf{v}|/|\mathbf{a}|\) where \(\mathbf{v}\) is the velocity and \(\mathbf{a}\) is the acceleration. \item The work time scale: \(\Phi/|\mathbf{v}\cdot\mathbf{a}|\) where \(\Phi\) is the gravitational potential, chosen to be 0 at large distances from the centre of the particle distribution. \item The escape time scale: $\sqrt{\Phi/\mathbf{a}\cdot\mathbf{a}}$. \end{enumerate} Each of the timesteps may be tuned with a dedicated prefactor, $\epsilon$, to reach the desired number of steps in an orbital period, $r$. Typically, this results in values of $\epsilon\approx0.01$. One may also disable an individual timestep criteria in practice by specifying a large prefactor. The timestep criteria are heuristics that ensure an individual particle achieves the desired precision in conservation of energy and momentum.

Particles at different phase-space locations in the simulation require significantly different timesteps. Therefore, we employ a multistep scheme (sometimes referred to as `block-step') based on a binary timestep tree to efficiently spend computational resources on particles that require smaller timesteps to achieve our required accuracy. A binary timestep tree can dramatically increase throughput, especially for the generic fully parallelized implementation in {\sc exp}. We begin by partitioning phase space $p$ ways such that each partition contains $n_j$ particles that require a timestep $\delta t=2^{-j} h$ where $h$ is the master time step and $j=0,\ldots,p$. Since the total cost of a time step is proportional to the number of force evaluations, the speed up factor is \begin{equation} \mathcal{S} = \sum_{j=0}^{p}n_j/\sum_{j=0}^{p} n_j 2^{-j}. \end{equation} We select the master timestep $h$ and the number of levels (sometimes referred to as `rungs') $p$ to spread out the particles among the levels.  The optimal value of $p$ depends on the range of frequencies in the simulation. For example, for an \(c=15\) NFW dark-matter profile with $N=10^7$ particles, we find that $p=7$ and \(\mathcal{S}\approx 30\).

\subsection{Coefficient interpolation}

The inclusion of multiple timesteps requires an additional scheme to correctly compute forces when only a fraction of the particles are advanced. Forces in the BFE algorithm depend on the expansion coefficients and the leap frog algorithm requires a linear interpolation of these coefficients to maintain second-order error accuracy per step.  This interpolation and the bookkeeping required for successive bisection of the time interval is straightforward.  We checked the accuracy of this algorithm by comparing it to direct orbit integration methods.

We refer to the set of coefficients that correspond to the contribution of particles at each individual multistep levels as the \emph{coefficient tableau}.  When computing the total coefficients at a particular multistep level, the offset of the velocity update at the half step in the kick-drift-kick leapfrog scheme allows the coefficient contribution at the lower inactive levels to be linearly interpolated.  The error in the contribution from the interpolation is the same order as that for leap frog itself.

Each particle is assigned a time step level, indexed as $j=0,\ldots,p$, based on the timestep criteria in the previous section. For each level with index $j$, the coefficient tableau is defined as the particles' contributions restricted to that level. We define an indicator $\zeta^j_i = 1$ if particle $i$ is in level $j$, and zero otherwise.  With this definition, the coefficient tableau for each function $\mu$ at each level $j$ becomes \begin{equation} \hat{a}_\mu^j(t+q/2^{p}) = \sum_{i=1}^N m_i \zeta_i^j \phi_\mu(x_i) \label{eq:coeft} \end{equation} where $q\in[0, 1, \ldots, 2^{p}-1]$ are the substeps. For particles at level $j$, each time substep is $h/2^j$.  We get the full coefficient by summing over $j$: \begin{equation} \hat{a}_\mu(t) = \sum_{j=0}^p \hat{a}_\mu^j(t).\end{equation}

To advance particles, we use the following procedure. For each sub-step $q$ (of $2^p$ total substeps), where particles at level $o\ge j$ satisfy $\zeta^j_i = 1$: \begin{enumerate} \item Define the fraction of the full step $h$: $f \equiv q/2^{p}$. \item For all levels $o\ge j$, compute the coefficient tableau $\hat{a}_\mu^o(t+q/2^{p})$. \item Compute the preceding step, $g_- \equiv \lfloor q/2^{p-o} \rfloor /2^{o}$, and the following step, $g_+ \equiv \lceil q/2^{p-o} \rceil /2^{o}$, where $\lfloor\cdot\rfloor$ and $\lceil\cdot\rceil$ are the floor and ceiling functions, respectively.  \item For all levels $o<j$ (i.e. $\zeta^j_i = 0$), interpolate the coefficient tableau using the preceding and following fractional steps: \begin{equation}\begin{aligned} \hat{a}_\mu^o(t+hf) = \frac{\hat{a}_\mu^o(t+hg_-) [f - g_-] + \hat{a}_\mu^o(t+hg_+) [g_+ - f]}{g_+ - g_-}. \\~ \end{aligned} \label{eq:cinterp} \end{equation} \item Advance all particles using the interpolated coefficients. \item Compute the new timestep for all particles and assign them to timestep levels. If a particle has changed levels, subtract the previous contribution from its former level and add its updated contribution to its new level. \end{enumerate}

For example, let us consider only two levels in total, i.e. $p=1$. Particles at level $j=0$ have time step $h$, the master timestep. Applying the kick and drift steps from the kick-drift-kick algorithm brings the positions of level 0 particles to the next timestep.  This allows us to evaluate \(\hat{a}_\mu^0(t)\), but to perform the final kick step, we also need the contribution from the $j=1$ particles: $\hat{a}_\mu(t+h) = \hat{a}_\mu^0(t+h) + \hat{a}_\mu^1(t+h)$.  The second term requires advancing the particles at higher levels, i.e. smaller timesteps. Now consider the level 1 particles.  The first substep brings the positions of level 1 particles to \(t + h/2\). To evaluate \(\hat{a}_\mu(t+h/2) = \hat{a}_\mu^0(t+h/2) + \hat{a}_\mu^1(t+h/2)\), \(\hat{a}_\mu^1(t+h/2)\) can be evaluated from the current positions, but we also need \(\hat{a}_\mu^0(t+h/2)\).  At time $t+h/2$, we have $q=1$. For $o=0$, we find $g_- = 0$ and $g_+ = 1$. Thus, using the linear interpolation formula (eq~\ref{eq:cinterp}): $\hat{a}_\mu^0(t+h/2) = \frac{\hat{a}_\mu^0(t) + \hat{a}_\mu^0(t+h)}{2}.$ This allows us to compute the next kick and drift, bringing all particles to the time $t+h$ that allows the evaluation of the final kick for all particles. The error in the force (acceleration) interpolation is \(\mathcal{O}(h)\). Propagating this error through the algorithm contributes to an error in the trajectory proportional \(h^3|\mathbf{x}^{[3]}|\). This is the same order as the leapfrog integrator itself so there is no need for a higher-order interpolation. A higher order symplectic integrator would require high-order interpolations.

\section{Summary and Conclusion}\label{sec:expsummary}

There are a number of reasons to use {\sc exp} to study galactic dynamics. First, it is efficient. The computational effort with BFE techniques scales as $\mathcal{O}\propto N$, rather than $\mathcal{O}\propto N^2$ for direct techniques or $\mathcal{O}\propto N\log(N)$ for tree-based techniques. The computational efficiency means that studies can push to higher $N$ when compared to other techniques. Owing to these computational considerations, in practice direct summation is not viable for most problems today. Additionally, because BFE does not require repeated domain decompositions to maintain efficiency, the code is easily ported to operate on graphical processing units (GPUs).

Second, while {\sc exp} requires a careful construction of the basis used during integration, once the basis is specified, the variance in forces at a given location in the model is low (see Sections~\ref{subsec:spherical} and \ref{subsec:cylindrical}). Thus, {\sc exp} is well-suited to study long-term secular evolution and subtle dynamical processes. For example, to study the subtle dynamical process of secular evolution in galaxy disks, like bar formation and evolution, one requires a gravity solver that is both accurate and low noise or the dynamical details can be compromised; orbits could switch families for numerical rather than for physical reasons, among other possible problems. Given the tests presented above, we feel that our {\sc exp} code is well suited for the study of bars, for example in \citet{petersen19} and \citet{petersen21}. In contrast, tree-based codes have larger force errors and force error variances.

Third, {\sc exp} enables after-the-fact studies such as in \citet{petersen21}. It is straightforward to extract a representation of the potential at every position in the model, which one may then use for detailed integration experiments. The BFE representation allows for isolation of different evolutionary modes, and so in a bar-like model, one may choose to rotate only the even modes, and further avoid components of the potential that are not related to the bar.

The primary disadvantage of {\sc exp} is that a BFE code is not fully adaptive, and thus cannot integrate arbitrary systems. For example, one should probably not use a BFE code like {\sc exp} if one wants to study small self-gravitating regions or complex geometries such as major mergers. In such situations a tree gravity code would probably be best.  The obvious bias pattern resulting from the basis in both the exponential disc and bar-and-disc case is also a BFE artefact that one must carefully study to ensure that the evolution is not appreciably changed (e.g. using coefficient significance analysis). No \(N\)-body gravity solver is a panacea, and one should endeavour to understand the advantages and limitations of any solver. 

The $n$-body solver {\sc exp} has been used to great effect in the literature to uncover subtle dynamical effects \citep[e.g.][]{weinberg02,holleyb05,weinberg07a,weinberg07b,choi07,choi09,petersen16a,petersen16b,petersen19,petersen21}. The {\sc exp} code is being prepared for a public release with an accompanying user manual describing the software design. We expect the highly accurate adaptive basis techniques in this work to not only continue producing valuable \(N\)-body models, but to also complement analytic work in dynamics as well as observational data studies.

\section*{Acknowledgements} This project made use of {\it numpy} \citep{numpy} and {\it matplotlib} \citep{matplotlib}. MSP acknowledges funding from a UK Science and Technology Facilities Council (STFC) Consolidated Grant.

\section*{Data Availability}
The data underlying this article will be shared on reasonable request to the corresponding author.

\bibliography{PetersenMS}

\end{document}